\documentclass[journal]{IEEEtran}
\usepackage{graphicx}
\usepackage{amsmath,amsfonts}
\usepackage{algorithmic}
\usepackage{algorithm}
\usepackage{array}
\usepackage{textcomp}
\usepackage{stfloats}
\usepackage{url}
\usepackage{verbatim}
\usepackage{graphicx} 
\usepackage{booktabs}
\usepackage{color}
\usepackage{multirow}
\usepackage{hyperref} 

\usepackage{graphicx}
\usepackage{subfigure}

\ifCLASSINFOpdf
\else
\fi
\begin{document}
%
\title{Single-cell Multi-view Clustering via Community Detection with Unknown Number of Clusters}
%
%
%

\author{Dayu~Hu, Zhibin~Dong,
        Ke~Liang, Jun~Wang, Siwei~Wang
        and~Xinwang~Liu$^{\dagger}$,~\IEEEmembership{Senior~Member,~IEEE}

\thanks{Dayu Hu, Zhibin Dong, Ke Liang, Jun Wang, and Xinwang Liu are with the School of Computer, National University of Defense Technology, Changsha, China, 410073. Email: hzauhdy@gmail.com, dzb20@nudt.edu.cn, liangke200694@126.com, wang\_jun@nudt.edu.cn, xinwangliu@nudt.edu.cn.}
\thanks{S. Wang is with Intelligent Game and Decision Lab,  Beijing 100071, China (e-mail: wangsiwei13@nudt.edu.cn.)}
\thanks{$^{\dagger}$ Corresponding author.}

}
 

%
%

\markboth{Journal of \LaTeX\ Class Files,~Vol.~14, No.~8, August~2015}%
{Shell \MakeLowercase{\textit{et al.}}: Bare Demo of IEEEtran.cls for IEEE Journals}
%



\maketitle

\begin{abstract}
Single-cell multi-view clustering enables the exploration of cellular heterogeneity within the same cell from different views. Despite the development of several multi-view clustering methods, two primary challenges persist. Firstly, most existing methods treat the information from both single-cell RNA (scRNA) and single-cell Assay of Transposase Accessible Chromatin (scATAC) views as equally significant, overlooking the substantial disparity in data richness between the two views. This oversight frequently leads to a degradation in overall performance. Additionally,  the majority of clustering methods necessitate manual specification of the number of clusters by users. However, for biologists dealing with cell data, precisely determining the number of distinct cell types poses a formidable challenge. To this end, we introduce scUNC, an innovative multi-view clustering approach tailored for single-cell data, which seamlessly integrates information from different views without the need for a predefined number of clusters. The scUNC method comprises several steps: initially, it employs a cross-view fusion network to create an effective embedding, which is then utilized to generate initial clusters via community detection. Subsequently, the clusters are automatically merged and optimized until no further clusters can be merged. We conducted a comprehensive evaluation of scUNC using three distinct single-cell datasets. The results underscored that scUNC outperforms the other baseline methods. 
\end{abstract}


%
\IEEEpeerreviewmaketitle

\section{Introduction}

\IEEEPARstart{S}{ingle-cell} sequencing technology embodies a state-of-the-art (SOTA) approach to high-throughput genome, transcriptome, and epigenome analysis at the individual cell level \cite{vanella2022high,shiau2023high,esain2023deciphering}. This advanced technology has emerged as a response to the limitations of traditional bulk sequencing \cite{kader2022evaluating}, enabling the exploration of gene modules at a heightened resolution. It assumes a pivotal role in elucidating the origins of tumors and their microenvironments, offering profound insights into cellular functions, developmental processes, and the mechanisms underlying diseases \cite{schreibing2022mapping}. In 2013, the prestigious journal Science acknowledged single-cell sequencing technology as one of the six most consequential areas in the realm of science. In 2015, it graced the cover of Science Translational Medicine. Presently, single-cell sequencing technology occupies a prominent position in various fields, including tumor research and biology, and has garnered increasing attention in life science research, with promising applications on the horizon \cite{gawad2016single,heo1989biology}. However, the recognition of distinct cell subpopulations presents challenges, as accurate cell assignment remains a complex issue \cite{hu2016single,rotem2015single}. While numerous cell labeling tools have been produced, they often grapple with achieving sufficient accuracy and rely on established ground truth \cite{pasquini2021automated,ji2023scannotate,cao2020scsa}. Manual annotation methods, characterized by their sluggishness and labor-intensiveness, make unsupervised clustering the preferred approach for cell allocation. Concurrently, fueled by advancements in sequencing technology, the joint analysis of multiple data sources has emerged as a prevailing trend \cite{stuart2019integrative,vandereyken2023methods,wen2022graph}. Single-cell clustering algorithms grounded in multiple views now represent the hot spot in research.

As deep learning has advanced, there have been notable strides in deep embedding learning within the realm of single-cell analysis \cite{li2022deep,yuan2022scbasset,cao2022multi}. Deep clustering techniques, particularly autoencoders (AE), have demonstrated their effectiveness in generating cellular representations. Numerous deep embedding methods rooted in AE have been introduced in prior literature \cite{li2022geometry,wang2022adversarial,gong2022deep,zhang2022embedding,sadeghi2023deep}. Furthermore, these methods have more recently been extended to address multi-view scenarios, which will be expounded upon in Sections II-A and II-B. 

\begin{figure}[!t]%
\centering
\includegraphics[width=0.8\linewidth]{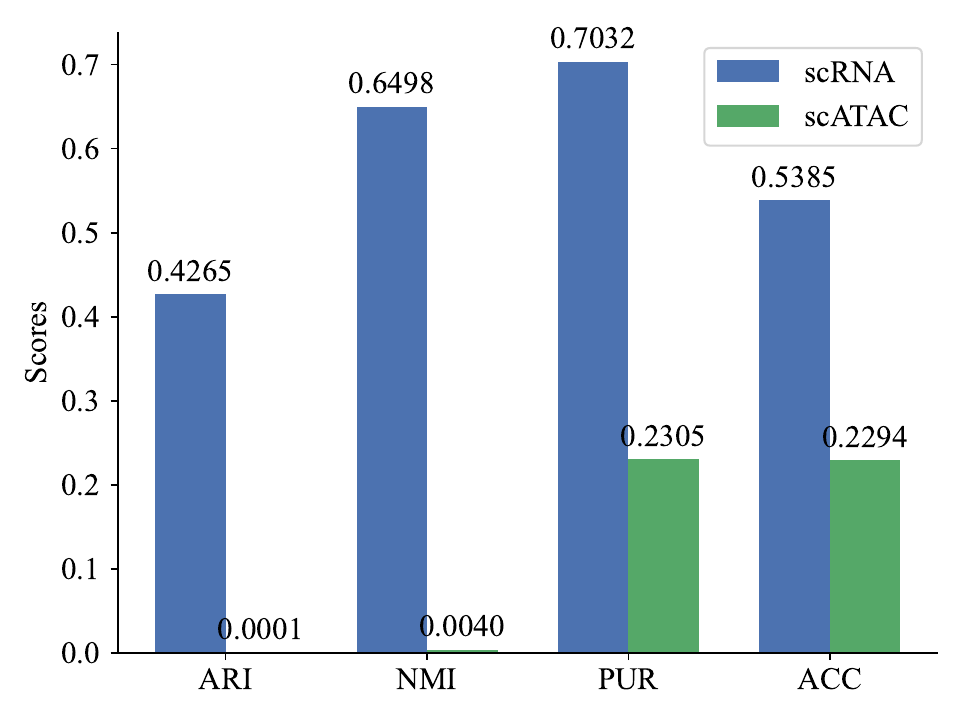}
\caption{
The $k$-means results for both scRNA and scATAC views highlight that scRNA view has a higher degree of information richness compared to scATAC.}\label{fig1}
\end{figure}

Nevertheless, this expansion poses a challenge to current research. Most existing multi-view clustering (MVC) algorithms crafted for single-cell data grapple with two primary issues.
The first challenge revolves around the underutilization of information across views. Differing from conventional multi-view datasets, single-cell multi-view data not only exhibit high dimensionality and sparsity but also encompass substantial information richness among distinct views \cite{kopp2022simultaneous}. As depicted in Fig. \ref{fig1}, the clustering performance on the scRNA view significantly outperforms that of the scATAC view. This disparity suggests that the scRNA view contains more abundant information compared to the scATAC view. Without proper treatment, both the two views are treated as equal, and the sparse scATAC information can detrimentally affect the final clustering performance. 

The second challenge revolves around the fact that nearly all existing MVC methods for single-cell data require users to manually specify the number of clusters. However, accurately determining the accurate number of clusters remains a formidable task for users. In a recent publication in Nature Methods, Grabski et al. emphasized that many existing single-cell clustering methods tend to exhibit over-clustering \cite{grabski2023significance}. Moreover, universal MVC methods often overlook cell-to-cell relationships, leading to suboptimal clustering performance on single-cell datasets. These common MVC methods with unknown number of clusters K (No-K) will be further discussed in Section II-C. While these methods have made commendable progress, they continue to grapple with effectively addressing the inherent complexities of single-cell multi-view data.

Considering all the factors mentioned earlier, we introduce an effective MVC framework for single-cell data, named scUNC. To address the disparity in information richness between the scRNA and scATAC views, we have devised a cross-view fusion network (CVFN), which automatically allocates weights to the scRNA and scATAC views. 
Furthermore, to deal with unknown number of clusters, we employ community detection on the derived cell representations to generate initial clusters. Through an iterative process, clusters are gracefully merged into larger clusters employing dip-tests, culminating in the attainment of convergence. The proposed scUNC method outperforms existing baseline methods on three real-world single-cell datasets.

In summary, the contributions of our paper can be outlined as follows:

\begin{itemize}
    \item We propose an effective MVC approach (scUNC) tailored for single-cell data with an unknown number of clusters. By introducing cross-view fusion, community detection and dip-test to handle unbalanced information richness of different views and unknown cluster numbers, scUNC is a pioneering work that integrates cross-view fusion module and No-K clustering into a joint framework.

    \item Cross-view fusion network is designed to allocate weights to both the scRNA and scATAC views, facilitating the generation of highly effective shared embeddings.

    \item Community detection is employed to establish the initial clusters, while the dip-test is utilized to iteratively merge these clusters until they converge. This procedure obviates the necessity for users to manually define the number of clusters.

    \item  Extensive experiments on three cellar benchmark datasets demonstrate the effectiveness of our model in terms of clustering performance.

\end{itemize}

\begin{figure*}[!t]%
\centering
\includegraphics[width=1\linewidth]{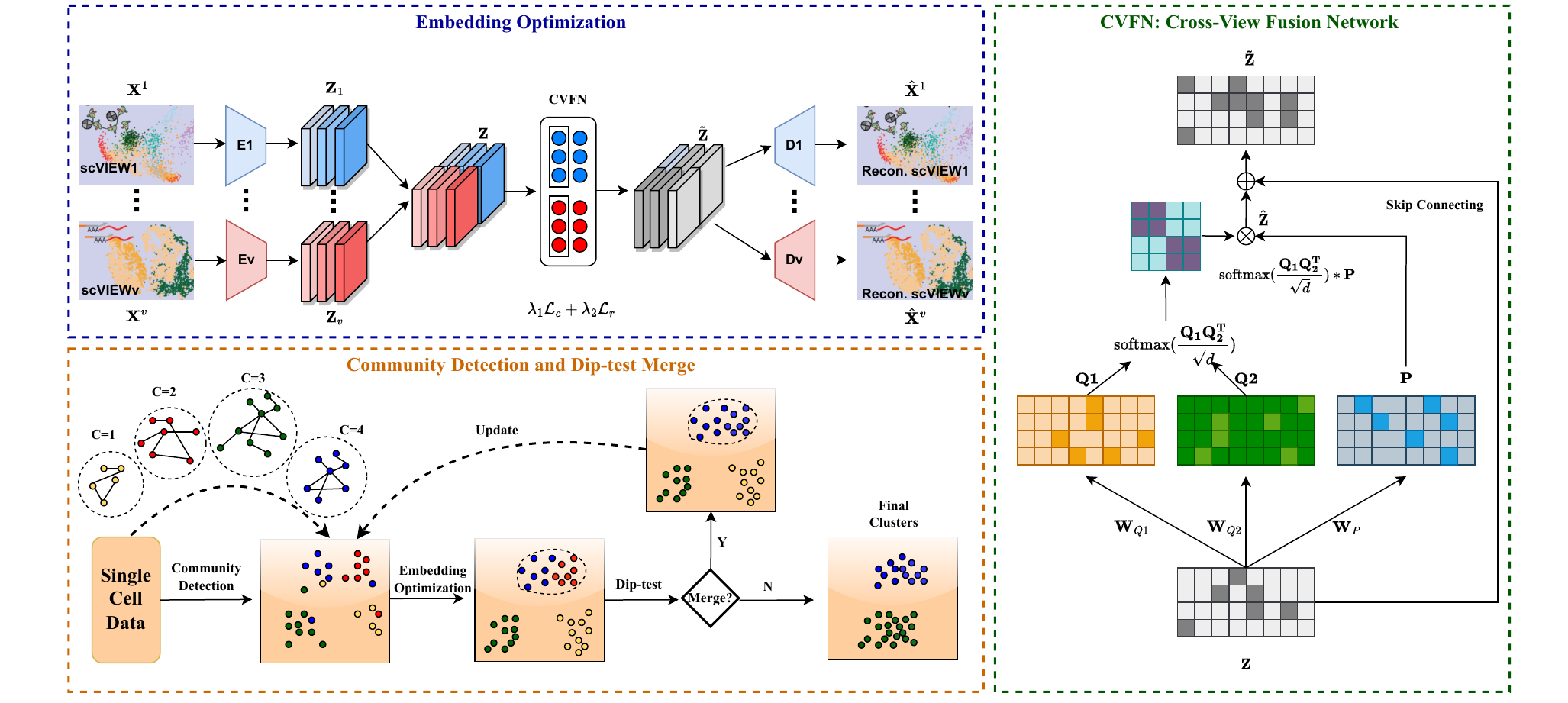}
\caption{The proposed framework scUNC encompasses three modules. (1) The CVFN module entails the mapping of the raw input into three distinct feature spaces, followed by the construction of a comprehensive structural relationship matrix, which enables the automated fusion of cross-view information. (2) The embedding optimization module combines reconstruction loss and clustering loss to jointly optimize the embeddings. (3) The community detection and dip-test merge module initially creates initial clusters using community detection methods and subsequently employs a dip-test for cluster merging until no further clusters can be merged, culminating in the output of the final clustering results. The optimization process and the automatic merging process operate in an alternating manner, mutually reinforcing each other.}\label{fig2}
\end{figure*}

\section{Related Work}

The existing algorithms designed for single-cell clustering can be classified into two types: single-view and multi-view methods. Additionally, we have also introduced some general MVC No-K clustering methods for comparison. In the subsequent text, a detailed introduction is provided.

\subsection{Single-view clustering methods for single-cell data}

Single-cell clustering within a single view has undergone extensive development over the years \cite{li2023single,zhang2022new,hu2023scdfc}. Early researchers experimented with base clustering models for single-cell data, including techniques like $k$-means and spectral clustering \cite{vzurauskiene2016pcareduce,cheng2018machine}. However, these models often struggled to capture the nonlinear characteristics inherent in cellular data. As research advanced, investigators turned to neural networks to extract deep features, resulting in the emergence of several autoencoder-based deep clustering methods. One notable example is DESC \cite{li2020deep}, which utilizes a neural network to learn a meaningful representation while effectively mitigating batch effects. Another significant contribution is scDeepCluster \cite{tian2019clustering}, where Tian et al. introduced a method that leverages the zero-inflated negative binomial (ZINB) and introduced a dedicated ZINB loss function for single-cell clustering. With the evolution of deep embedding techniques, some researchers, drawing inspiration from cell interactions, explored the potential of deep graph embedding clustering methods. Cheng et al. introduced an attention-based single-cell graph clustering algorithm \cite{cheng2022scgac}, while Gan et al. proposed a deep structural clustering framework that adeptly preserves both the graph structure information between cells and the node attribute data of the cells themselves \cite{gan2022deep}. In summary, while there has been notable progress, the increasing multi-view data poses a challenge for single-view-based single-cell clustering methods. These methods encounter limitations in their capacity to effectively integrate cross-view information.

\subsection{Multi-view clustering methods for single-cell data}

Single-cell multi-view data has emerged in recent years, driven by advancements in sequencing technology that enable researchers to collect information from various views of the same cell \cite{wang2023multi}. The integration of data from multiple views provides a comprehensive understanding of cell characterization. Zuo et al. pioneered the development of scMVAE \cite{zuo2021deep}, a neural network based on variational autoencoders, explicitly designed for feature extraction from both scRNA and scATAC data. This model employs a Gaussian probability model to approximate the data distribution of cell matrices and has demonstrated promising results, while this premise does not consistently hold true. In addition, the DCCA model leverages data from one omics to fine-tune data from another, effectively amalgamating information from different views \cite{zuo2021deepcross}. More recently, Ren et al. introduced a multi-view approach based on subspace clustering \cite{ren2023scmcs}, aimed at reducing information redundancy between subspaces to generate high-quality cell representations.

\subsection{Multi-view clustering methods with unknown number of clusters}

Here, we will introduce the MVC method with an unknown number of clusters. These techniques are considered general approaches and are not specifically tailored for single-cell data. The early and well-known work in general multi-view No-K clustering is COMIC \cite{peng2019comic}. This method learns a connection graph within a projection space while concurrently minimizing dissimilarity between pairwise connection graphs derived from different views. In addition, RST-MVC develops a novel regularization technique to autonomously obtain the cluster number from the underlying data distribution \cite{yuan2022robust}. While these methods have exhibited commendable performance in various contexts, they may not ideally accommodate the data distribution of single-cell data due to the presence of intricate relationships between cells. There are inherent cell communities among cells, but these methods may not effectively recognize these communities.

\begin{table}[t]
\centering
\renewcommand{\arraystretch}{1.5}
\caption{Notation summary}
\vspace{-0.2 cm}

\begin{tabular}{cc}
\hline
\textbf{Notation} & \textbf{Explanation}
\\\hline

$\mathbf{X}^{1},\mathbf{X}^{v}$& Input data for the $1$-th and the $v$-th view.\\

$\mathbf{Z}^{1},\mathbf{Z}^{v}$&  Encoded embedding for the $1$-th and the $v$-th view.\\
$\mathbf{Z},\hat{\bf{Z}}$& Embedding after concatenation and fusion.\\

$\tilde{\bf{Z}}$& Embedding outputed via CVFN network.\\

$\mathbf{Q}_1,\mathbf{Q}_2$& Embeddings used to compute structural relationship.\\

$\mathbf{P}$& Embeddings used to map the input.\\

$\mathbf{W}_{ij}$& Weights between node $i$ and node $j$.\\

$\mathbf{Q}_c$& The modularity coefficient.\\

$\mathcal{P}_{\mathrm{dip}}(\cdot)$& The dip-score of dip-test.\\

$\mathbf{C}$& The community exists among cells.\\

$\mathcal{L}_{\mathrm{r}},\mathcal{L}_{\mathrm{c}},\mathcal{L}_{\mathrm{f}}$&  Reconstruction, clustering and total loss.\\

$\lambda_{1},\lambda_{2}$&  Two hyperparameters that balance the losses.\\

$\hat{\mathbf{X}}^{1},\hat{\mathbf{X}}^{v}$&  Reconstructed data  for the $1$-th and the $v$-th view.\\

\hline
\end{tabular}

\vspace{-0.2 cm}
\label{table1} 
\end{table}

\section{Methods}
\subsection{Preliminary}

Single-cell multi-view data pertains to the complex biological multi-modal data acquired through single-cell sequencing techniques. With the continuous advancement of technology, there is a growing interest in exploring multiple biological attributes. In this work, we provide a simple mathematical description of this cell problem. The datasets are represented as a multi-view matrix, denoted as  $\{\mathbf{X}^{v}=\{\mathbf{x}_{1}^{v};...;\mathbf{x}_{N}^{v}\}\in \mathbb{R}^{N \times M_{v}}\}_{v=1}^V$, where $\mathbf{X}^v$ corresponds to the data from the $v$-th view, $M_v$ denotes the dimension of genes from the $v$-th view. $V$ represents the count of views, while N represents the number of cells.


To enhance clarity and facilitate comprehension, given the extensive use of notations in this paper, we have provided a comprehensive table of notations for reference, which is outlined in Table \ref{table1}.

\subsection{Overview}

Our proposed scUNC framework is displayed in Fig. \ref{fig2}. One of the key advantages of this framework is its ability to assign optimal weights to scRNA and scATAC and effectively fuse them. Another advantage is that our framework eliminates the need for manual specification of the number of clusters, which is particularly beneficial for biologists conducting cell cluster analysis.

To elaborate further, following established practices, we begin by excluding outlier cells. Subsequently, we employ multiple autoencoders, denoted as $\{\mathbf{E}^{v}\}_{v=1}^V$, to transform the original feature matrix, $\{\mathbf{X}^{v}\}_{v=1}^V$, into a series of low-dimensional representations designated as $\{\mathbf{Z}^{v}\}_{v=1}^V$, and the corresponding shared embedding is denoted as $\textbf{Z}$. In the absence of additional operations, the network would accord equal importance to information from both the scRNA and scATAC views. However, holding that the scATAC view inherently contains notably less information compared to the scRNA view, this equitable treatment might potentially impede the overall performance of the integrated cell representation, especially when dealing with views characterized by limited information richness. To rectify this concern, we have devised a CVFN network that autonomously assigns weights to individual perspectives based on their information richness, thereby effectively redressing this imbalance.

Moreover, different from the majority of existing frameworks that employ $k$-means on the low-dimensional cell representations to establish initial clustering centers, we adopt a distinctive approach, i.e., community detection to form initial clusters, since it adeptly captures the intercellular community that exists among cells. A comprehensive exposition of community detection will ensue in Section III-D. 

After acquiring the initial cluster centroids through community detection, our motivation is to merge these clusters.   Drawing inspiration from the dip-test\cite{hartigan1985dip}, a widely used statistical tool, we undertake a statistical assessment of the clusters in each iteration to ascertain their states for merging. This iterative merging process persists until convergence is achieved, culminating in the final clustering outcomes.

\subsection{Cross-View Fusion Network}

In this paper, we elucidate the substantial disparity in information richness among different views of single-cell data, with the scRNA view exhibiting significantly greater information compared to scATAC.  Consequently, it is challenging to effectively fuse information from multiple views.  Here, we have devised the Cross-View Fusion Network to automate the allocation of optimal weights to different views.

Subsequently,  we will delve into the structure of the CVFN network in Fig. \ref{fig2}. Starting with the original cell feature matrices $\{\mathbf{X}^{v}\}_{v=1}^V$, we first employ encoders to extract their low-dimensional representations, which are subsequently concatenated to yield a consolidated representation $\mathbf{Z}$, bearing the amalgamated information from all views.      This can be succinctly articulated as follows:
\begin{equation}
\label{eq:1}
\begin{aligned}
\mathbf{Z} =\left[\mathbf{Z}^{1}, \mathbf{Z}^{2}, \ldots, \mathbf{Z}^{v}\right].
\end{aligned}
\end{equation}

With the advent of the era of large-language models, the Transformer architecture has gained increasing recognition among researchers. Drawing inspiration from the Transformer \cite{liu2023survey,zhang2023self,kaselimi2022vision}, we have devised three feature transformations, denoted as $\mathbf{W}_{Q_1},\mathbf{W}_{Q_2}$ and $\mathbf{W}_{P}$, which map the shared representation $\mathbf{Z}$ into three distinct feature spaces, yielding three transformed embeddings, $\mathbf{Q}_{1},\mathbf{Q}_{2}$ and $\mathbf{P}$. The process of feature transformation can be mathematically articulated as follows:
\begin{equation}
\label{eq2}
\begin{aligned}
\mathbf{Q}_1 =\mathbf{Z}\mathbf{W}_{Q_1};
\mathbf{Q}_2 =\mathbf{Z}\mathbf{W}_{Q_2};
\mathbf{P} =\mathbf{Z}\mathbf{W}_{P}.
\end{aligned}
\end{equation}

We then utilize $\mathbf{Q}_{1}$ and $\mathbf{Q}_{2}$ to compute the matrix $\mathbf{S}$ representing the structural relationships between cells:
\begin{equation}
\label{eq:3}
\begin{aligned}
\mathbf{S}=\operatorname{softmax}\left(\frac{\mathbf{Q}_1\mathbf{Q}_2^{T}}{\sqrt{d}}\right), 
\end{aligned}
\end{equation}

\noindent where $d$ represents the input dimension of embedding $\mathbf{Z}$. Subsequently, we multiply the obtained structural relationship matrix $\mathbf{S}$ by the transformed representation $\mathbf{P}$, resulting in the embedding $\hat{\bf{Z}}$ enhanced by structural information. This process can be expressed by the following formula:
\begin{equation}
\label{eq:4}
\begin{aligned}
\hat{\bf{Z}} =\mathbf{S}\mathbf{P}.
\end{aligned}
\end{equation}

Finally, to prevent network degradation, we combine the input representation $\mathbf{Z}$ to the obtained $\hat{\bf{Z}}$ via skip connections. This transforms the original feature transformation process into a fine-tuning process for the embedding $\mathbf{Z}$. The final output to the CVFN network denoted as $\tilde{\bf{Z}}$, is derived from the combination of $\mathbf{Z}$ and $\hat{\bf{Z}}$, this process can be denoted as:
\begin{equation}
\label{eq:5}
\begin{array}{l}
\tilde{\bf{Z}} = {\mathbf{W}_{\mathbf{Z}}}\left( \bf{Z}+\hat{\bf{Z}}\right) + {\mathbf{b}},
\end{array}
\end{equation}

\noindent where $\mathbf{W}_{\mathbf{Z}}$ refers to the learnable matrix, $\mathbf{b}$  denotes bias corresponding to the transformation via skip connections.

\subsection{Community Detection}
Community detection is a technique used to assign nodes to communities based on their neighbor relationships, commonly employed for analyzing networks \cite{berahmand2022graph,kazemzadeh2022influence,li2022modified,safdari2022reciprocity,li2022motif}. Given the inherent presence of extensive interaction networks between cells, it is particularly well-suited for analyzing single-cell data. The objective of community detection algorithms is to divide the network into multiple communities, such that connections between nodes within each community are more densely interconnected, while inter-community edges are relatively sparse. This serves the same function as the $k$-means algorithm, and as such, in this framework, we employ community detection in place of $k$-means to generate initial clustering centers of embeddings $\tilde{\bf{Z}}$.

In this section, we introduce the specific process of community detection within our framework. Initially, using the final embeddings $\tilde{\bf{Z}}$ resulting from the CVFN network, we identify k nearest neighbors for each cell node through a K-nearest neighbors (KNN) approach. Subsequently, we initialize each cell node as an independent community while simultaneously calculating the weights between different nodes, with the calculation formula being:
\begin{equation}
\label{eq:6}
\begin{aligned}
\mathbf{W}_{i j} = \frac{|\mathbf{v}(i) \cap \mathbf{v}(j)|}{|\mathbf{v}(i) \cup \mathbf{v}(j)|},
\end{aligned}
\end{equation}

\begin{algorithm}[tb]
\caption{Community Detection Algorithm}
\label{alg:louvain}
\small
\textbf{Input}: Embeddings $\tilde{\bf{Z}}$; Initial community assignment $\mathcal{C}$\\
\textbf{Output}:Community assignment $\mathcal{C}$
\begin{algorithmic}[1]
\REPEAT
\STATE $\text{modularity\_improved} \gets \text{False}$
\FOR{each node $v$ in $\tilde{\bf{Z}}$}
\STATE $\Delta Q_{\text{max}} \gets 0$
\STATE $\text{best\_community} \gets \text{current\_community}(v)$
\FOR{each community $c$ in the neighborhood of $v$}
\STATE $\Delta Q \gets \text{calculate\_modularity\_gain}(v, c)$
\IF{$\Delta Q > \Delta Q_{\text{max}}$}
\STATE $\Delta Q_{\text{max}} \gets \Delta Q$
\STATE $\text{best\_community} \gets c$
\ENDIF
\ENDFOR
\IF{$\Delta Q_{\text{max}} > 0$}
\STATE $\text{move\_node}(v, \text{best\_community})$
\STATE $\text{modularity\_improved} \gets \text{True}$
\ENDIF
\ENDFOR
\UNTIL{not $\text{modularity\_improved}$}
\end{algorithmic}
\end{algorithm}

\noindent where $\mathbf{v}(i)$ is the neighbor set of node $i$, and $\mathbf{v}(j)$ is the set of node $j$.  
For the sake of convenience in the following introduction, we give the concept of the modularity coefficient $\mathbf{Q}_{c}$, defined as follows:
\begin{equation}
\label{eq:7}
\begin{aligned}
\mathbf{Q}_{c}=\frac{1}{2m} \sum_{i, j}\left[\mathbf{W}_{i j}-\frac{\mathbf{s}_i \mathbf{s}_j}{2m}\right] \delta(\mathbf{c}_i, \mathbf{c}_j).
\end{aligned}
\end{equation}

\begin{figure}[t]
\centering
\includegraphics[width=0.7\columnwidth]{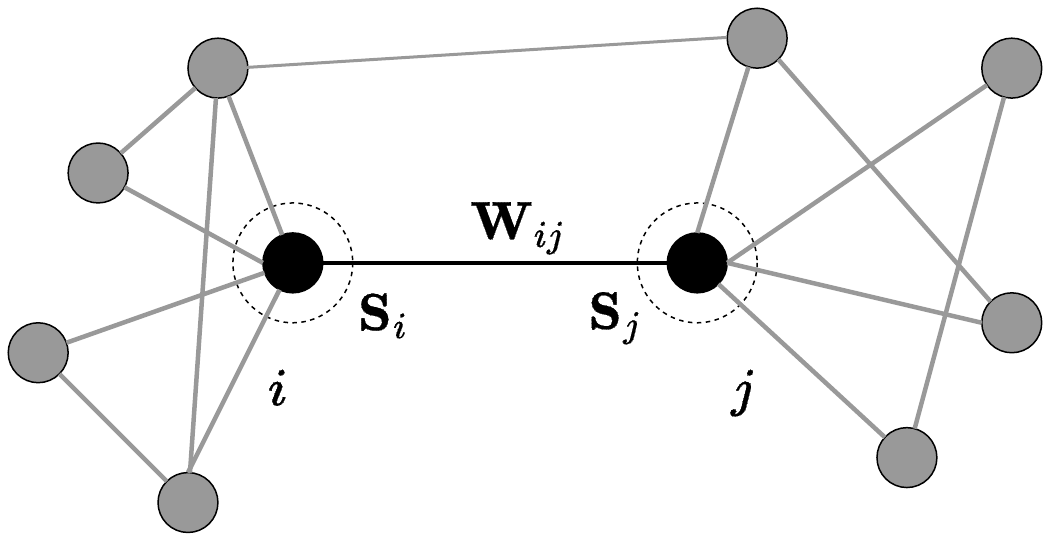} %
\vspace{-0.2 cm}
\caption{An illustrative representation of the contribution of two nodes to the calculation of modularity.}
\vspace{-0.2 cm}
\label{fig3}
\end{figure}

For a series of communities $\mathbf{C}=\{\mathbf{c}_{1};...;\mathbf{c}_{k}\}$, $\mathbf{c}_{i}$ represents the community to which node $i$ belongs, and similarly, $\mathbf{c}_{j}$  denotes the community to which node $j$ belongs. $\mathbf{W}_{i j}$ is employed to measure the weight of the   $i$-th and $j$-th node, while $\mathbf{s}_{i}$ and $\mathbf{s}_{j}$ represent the sum of edge weights between node $i$ and all other nodes, and node $j$ and all other nodes, respectively. $m$ represents a standardized constant. As presented in Fig. \ref{fig3}.

The Kronecker delta used in Formula \ref{eq:8}, denoted as $\delta(i, j)$, is written as follows:
\begin{equation}
\label{eq:8}
\begin{aligned}
\mathbf{\delta}(i, j) = \begin{cases}
1, & \text{if } i = j \\
0, & \text{if } i \neq j
\end{cases}.
\end{aligned}
\end{equation}

In this framework, the delta function is used to determine whether $i$-th and $j$-th nodes belong to the same community. Ultimately, the modularity coefficient $\mathbf{Q}_{c}$ has been computed, falling within the range of -1 to 1. This metric functions as an objective gauge of the partitioning quality for a network into clusters. Consequently, the problem is transmuted into a combinatorial optimization challenge.

Subsequently, we expound upon our method for the initial clustering assignment of the obtained embeddings $\tilde{\bf{Z}}$, which relies on the modularity coefficient. As delineated in Algorithm 1, we initiate the process by identifying the k-nearest neighbors of each node through the utilization of a KNN technique. Following this, for each node, we compute its modularity gain concerning nodes within various communities. The node is transferred to the target community that offers the maximum modularity gain. This iterative procedure persists until no additional modularity gains can be achieved. Ultimately, each cell node is assigned to distinct clusters.

\subsection{Dip-test}

The dip-test is a statistical test method introduced by Hartigan et al. in 1985 to measure the modality of a data distribution \cite{hartigan1985dip}.   This test examines sample sets extracted from two clusters, providing a dip-score as the output.  The dip-score signifies the probability of unimodality within the sample sets.   A higher dip-score indicates that the two clusters exhibit a high level of structural similarity.   Within our framework, the dip-score is utilized to determine whether clusters should be merged.   We hold that clusters with high structural similarity are considered to belong to the same category.

The computation process of the dip-test is rather intricate, involving the calculation of unimodal depth and total depth of the dataset. However, we won't delve into the details here. For the sake of simplicity, we denote the dip-test as  $\mathcal{P}_{\mathrm{dip}}(\cdot)$.

\subsection{Overall Optimization Module}

In this section, we will present a comprehensive optimization module offered by scUNC, which includes the reconstruction loss, clustering loss, and merging process.
\subsubsection{Reconstruction loss}

The scUNC model is constructed upon an autoencoder, an unsupervised learning algorithm that aims to retain the essential features of input data and create an efficient representation of the original data. Our initial process aligns with that of many existing frameworks, involving the compression of original information into a bottleneck representation, followed by its reconstruction using the bottleneck layer by decoder $\{\mathbf{D}^{v}\}_{v=1}^V$. Represented as the $v$-th view, the reconstructed information can be expressed as follows:
\begin{equation}
\label{eq:9}
\begin{aligned}
\hat{\mathbf{X}}^{v} = \tilde{\bf{Z}}\mathbf{W}^{v}_D,
\end{aligned}
\end{equation}

\noindent here, $\mathbf{W}^{v}_D$ represents a learnable weight matrix used for decoding in the $v$-th view. Its function is to remap the representations $\tilde{\bf{Z}}$ obtained through the previous CVFN network back to the original feature space. To enhance the quality of the generated representations, we introduce a reconstruction loss, which measures the 2-norm loss between the reconstructed data and the input data. Since the data in this paper consists of multiple cell views, the reconstruction loss is the cumulative sum of the reconstruction losses for each view. This process can be represented as follows:
\begin{equation}
\label{eq:10}
\begin{aligned}
\mathcal{L}_{\mathrm{r}}=\sum_{v=1}^{V}\left\|\mathbf{X}^{v}-\hat{\mathbf{X}}^{v}\right\|_{2}^{2}.\\
\end{aligned}
\end{equation}

\begin{algorithm}[tb]
\caption{Automatic Clustering Algorithm}
\label{alg:algorithm}
\small
\textbf{Input}: Embedding $\mathbf{Z}$; The number of epochs $t$  \\
\textbf{Output}: Final clustering assignment $\mathcal{\hat{C}}$

\begin{algorithmic}[1] 
\STATE Initialize clustering form $\mathbf{Z}$ via community detection: $\mathcal{\hat{C}}$
\FOR{t epochs}
\STATE $\text{Optimization through reconstruction loss via Eq.(10)}$.
\STATE $\text{Optimization through clustering loss via Eq.(11)}$.
\REPEAT
\STATE Select two clusters to calculate the Dip-score: 

\STATE $\text{Dip-score} = \text{dip-test}(\text{Cluster}^{i}, \text{Cluster}^{j})$
\STATE Merge clusters if $\text{Dip-score}>\text{threshold}$: 
\STATE$\text{Cluster}^{new} = \mathrm{MergeClusters}(\text{Cluster}^{i}, \text{Cluster}^{j})$

\STATE$\text{Update}$ $\mathcal{\hat{C}}$

\UNTIL{{No clusters can be merged}}
\ENDFOR
\end{algorithmic}
\end{algorithm}

\subsubsection{Clustering loss}
Relying solely on the reconstruction loss is not sufficient to impose enough constraints on the cell representations. Therefore, we introduce a clustering loss to facilitate joint optimization, which is defined as follows:
\begin{equation}
\label{eq:11}
\begin{aligned}
\mathcal{L}_{\mathrm{c}}=\frac{\left(1+\sigma_{\textbf{D}_{c}}\right)}{\left(\bar{\textbf{D}}_{c}\right)}   \frac{1}{|N|} \sum_{j=1}^{N} \sum_{i=1}^{K} \mathcal{\hat{P}}_{\mathrm{dip}}\left(c_{j}, i\right) \| \tilde{\bf{z}_{j}}-\mu_{i} \|_{2}^{2},
\end{aligned}
\end{equation}

\noindent here, $\textbf{D}_{c}$ represents cluster-pairwise distances, $\bar{\textbf{D}}_{c}$ represents the mean of $\textbf{D}_{c}$, and $\sigma_{\textbf{D}_{c}}$ represents the standard deviation of $\textbf{D}_{c}$. $\mu$ denotes the centroids of K clusters. $\mathcal{\hat{P}}_{\mathrm{dip}}(\cdot)$ represents the normalized dip-test $\mathcal{P}_{\mathrm{dip}}(\cdot)$ , which reflects the similarity between clusters. The normalization process is as follows:
\begin{equation}
\label{eq:12}
\begin{aligned}
\mathcal{\hat{P}}_{\mathrm{dip}}(c_{j}, i) = \frac{\mathcal{P}_{\mathrm{dip}}(c_{j}, i)}{\sum_{t=1}^{K} \mathcal{P}_{\mathrm{dip}}(c_{j}, t)}.
\end{aligned}
\end{equation}

Furthermore, the cluster-pairwise distances, denoted as $\textbf{D}_{c}$, are defined as follows:
\begin{equation}
\label{eq:13}
\begin{aligned}
\textbf{D}_{c}=\left\{\sqrt{\left\|\mu_{i}-\mu_{t}\right\|_{2}^{2}} \mid i \in[1, K-1] \text { and } t \in[i+1, K]\right\}.
\end{aligned}
\end{equation}

In essence, our model refines the embeddings by minimizing the cell representation's disparity from the assigned cluster centers.   The $\mathcal{\hat{P}}_{\mathrm{dip}}(\cdot)$, serves as a scalar, reflecting heightened resemblances amidst clusters.   Consequently, throughout the optimization procedure, clusters exhibiting elevated dip-scores progressively converge.   This result concurs with the design principles of our workflow, iteratively merging similar clusters. Furthermore, we have integrated the $\textbf{D}_{c}$-based standard deviation to promise the scale simultaneously pulls single clusters to a distant position.

\subsubsection{Merging process}

In the previous section, we introduced the clustering loss and reconstruction loss. Ultimately, we perform joint optimization by incorporating both the clustering loss and the reconstruction loss. The final loss function L can be mathematically expressed as follows:
\begin{equation}
\label{eq:14}
\begin{aligned}
\mathcal{L}_{\mathrm{f}}= \lambda_{1}\mathcal{L}_{\mathrm{c}} + \lambda_{2}\mathcal{L}_{\mathrm{r}},
\end{aligned}
\end{equation}

\noindent where $\lambda_{1}$ and $\lambda_{2}$ are two hyperparameters used to balance the clustering loss and reconstruction loss. In the subsequent section, we will examine the parameter sensitivity of these two values.

Next, we introduce the complete clustering procedure, where the optimization module and the automated merging module collaborate concurrently, as illustrated in Algorithm 2. Upon acquiring the fused cell representation $\tilde{\bf{Z}}$ from the CVFN network, we employ a community detection algorithm to generate the initial clusters. Subsequently, we evaluate the obtained clusters based on the dip-test defined earlier in the text. Clusters with higher dip-scores are considered highly correlated and merged into the same cluster. Following the completion of merging, the optimization process continues while simultaneously conducting the dip-test evaluation for the next round of clusters. It is worth noting that in this workflow optimization and merging process are not isolated, they operate alternately and mutually reinforce each other. The algorithm iteratively runs until no clusters can be further merged, eventually producing the final allocation results of the clusters. In summary, our automatic clustering algorithm not only eliminates the need for manual parameter configuration but also automatically brings similar clusters closer together and merges them, resulting in high-quality clusters.

\section{Experiments}
We conducted comprehensive experiments on three distinct real single-cell multi-view datasets and a non-cell dataset across different scales. The following sections will be introduced by the order of experimental settings, clustering performance, ablation study, model analysis, and generalization analysis.

\begin{table*}[t]
\centering
\small
\renewcommand{\arraystretch}{1.5}
\caption{Clustering result comparison for three real-world single-cell datasets. The top two performers are highlighted in bold.}
\vspace{-0.2 cm}
\resizebox{\textwidth}{!}{
\begin{tabular}{c|c|c|c|c|c|c|c|c|c|c|c|c}
\hline
Datasets & \multicolumn{4}{c|}{BMNC} & \multicolumn{4}{c|}{SMAGE-10K} & \multicolumn{4}{c}{SMAGE-3K} \\
\hline
Metrics & ARI & NMI & PUR & ACC & ARI & NMI & PUR & ACC & ARI & NMI & PUR & ACC  \\
\hline
$k$-means & 0.5205 & 0.7443 & \textbf{0.8515} & 0.5565 & 0.465 & 0.5861 & 0.8309 & 0.5594 & 0.5109 & 0.5807 & 0.7667 & 0.5799  \\
Spectral & 0.4497 & 0.6919 & 0.5799 & 0.5015 & 0.4982 & 0.5679 & 0.8079 & {\textbf{0.6525}} & 0.5389 & 0.5989 & 0.7729 & 0.6267 \\
DESC & 0.5125 & 0.6872 & 0.6504 & 0.5435 & 0.3263 & 0.5322 & 0.7838 & 0.4509 & 0.5360  & 0.5664 & 0.6716 & 0.6395 \\
scDeepCluster & 0.5676 & 0.7572 & 0.8200 & 0.6383 & 0.3518 & 0.5604 & {\textbf{0.8432}} & 0.4794 & 0.3929 & 0.5740 & \textbf{0.7838} & 0.5721 \\
scDSC & 0.6193 & 0.6504 & 0.6820 & 0.6269 & \textbf{0.5102} & 0.5314 & 0.7426 & 0.6391 & \textbf{0.5514} & {\textbf{0.6189}} & 0.7574 & \textbf{0.6770} \\
DCCA & 0.4912 & 0.7277 & 0.6101 &\ 0.5816 & 0.3866 & 0.5511 & 0.5200  & 0.4648 & 0.2984 & 0.5473 & 0.4588 & 0.4847 \\
scMVAE & 0.4225 & 0.706 & 0.539 & 0.4950  & 0.3430  & 0.5726 & 0.5240  & 0.4888 & 0.3616 & 0.5794 & 0.5284 & 0.5234 \\
scMCs & 0.1841 & 0.3906 & 0.4722 & 0.2517 & 0.2471 & 0.3598 & 0.7428 & 0.3561 & 0.2505 & 0.4255 & 0.7087 & 0.3992 \\
scMDC & \textbf{0.7126} & \textbf{0.8049} & 0.7077 & \textbf{0.6704} & 0.4528 & \textbf{0.5951} & 0.8208 & 0.5798 & 0.4722 & 0.5975 &{\textbf{0.7884}} & 0.6016  \\
scUNC (Ours) & {\textbf{0.8406}} & {\textbf{0.8458}} & {\textbf{0.8632}} & {\textbf{0.8530}} & {\textbf{0.6067}} & {\textbf{0.6422}} & \textbf{0.8317} & \textbf{0.6516} & {\textbf{0.5672}} & \textbf{0.6090} & 0.6839 & {\textbf{0.6839}}  \\
\hline
\end{tabular}
}
\vspace{-0.2 cm}

\label{table2}
\end{table*}

\begin{table}[t]
\centering
\vspace{-0.2 cm}
\renewcommand{\arraystretch}{1.5}
\small
\caption{The summary of datasets}
\resizebox{\linewidth}{!}{
\begin{tabular}{ccccccccc} 
\hline
                & Dataset   & Samples  & View1 & View2&Clusters \\ 
\hline
\multirow{3}{*}{Single-cell} 
& BMNC &  30672 & 1000 & 25& 27\\                           
                             & SMAGE-10K &  11020 & 2000 & 2000& 12\\
                             & SMAGE-3K      & 2585 & 2000 & 2000& 14  \\
\hline
{Non-cell}          & WikipediaArticles & 693 & 128  & 10 & 10           \\

\hline

\hline                            
\end{tabular}}
\vspace{-0.2 cm}
\label{table3}

\end{table}

\subsection{Experimental Settings}

\subsubsection{Datasets}

In this work, a total of four multi-view datasets are included, with three of them being real single-cell datasets, and the remaining one is a non-cell dataset used for assessing the model's generalization ability. An overview of the dataset information is presented in Table \ref{table3}, including details about sample sizes, feature counts, and the number of clusters. A brief introduction to each dataset is provided below:

\begin{itemize}
\item \textbf{BMNC}: The BMNC dataset and its associated cell type labels have been acquired from the 'bmcite' dataset within the 'SeuratData' package\footnote{https://github.com/satijalab/seurat-data}.

\item 
\textbf{SMAGE-10K}: The SMAGE-10K dataset includes human peripheral blood mononuclear cells (PBMCs) with about 10k cells, which has been obtained directly from the official 10X Genomics website\footnote{https://www.10xgenomics.com/resources/datasets}.

\item 
\textbf{SMAGE-3K}: Similarly, the SMAGE-3K dataset includes human PBMCs with about 3k cells, which has been obtained directly from the official 10X Genomics website\footnotemark[2].

\item \textbf{WikipediaArticles}: This dataset comprises carefully chosen segments extracted from Wikipedia's distinguished featured articles collection\footnote{http://www.svcl.ucsd.edu/projects/crossmodal/}.

\end{itemize}

\subsubsection{Compared Methods}

Our scUNC method is tailored specifically for single-cell data. In this study, to assess its effectiveness, we compared it with nine clustering algorithms designed specifically for single-cell data. These methods can be categorized into three groups as below. 

Four single-cell multi-view clustering algorithms:

\begin{itemize}

\item 
\textbf{scMDC} \cite{lin2022clustering}: Clustering of single-cell multi-omics data with a multimodal deep learning method

\item \textbf{scMVAE} \cite{zuo2021deep}: Deep-joint-learning analysis model of single cell transcriptome and open chromatin accessibility data.

\item \textbf{scMCs} \cite{ren2023scmcs}: scMCs: a framework for single-cell multi-omics data integration and multiple clusterings.

\item \textbf{DCCA} \cite{zuo2021deepcross}: Deep cross-omics cycle attention model for joint analysis of single-cell multi-omics data.

\end{itemize}

Three single-cell single-view clustering algorithms:

\begin{itemize}

\item 
\textbf{DESC} \cite{li2020deep}: Deep learning enables accurate clustering with batch effect removal in single-cell RNA-seq analysis.

\item \textbf{scDeepCluster} \cite{tian2019clustering}: Clustering single-cell RNA-seq data with a model-based deep learning approach.

\item \textbf{scDSC} \cite{gan2022deep}: Deep structural clustering for single-cell RNA-seq data jointly through autoencoder and graph neural network.

\end{itemize}

Two basic clustering methods:

\begin{itemize}

\item 
\textbf{K-means} \cite{hartigan1979algorithm}: Algorithm AS 136: A $k$-means clustering algorithm.

\item \textbf{Spectral clustering} \cite{von2007tutorial}: A tutorial on spectral clustering.

\end{itemize}

\begin{figure*}[htbp]
    \centering
    \subfigure[]{\includegraphics[width=1.5in]{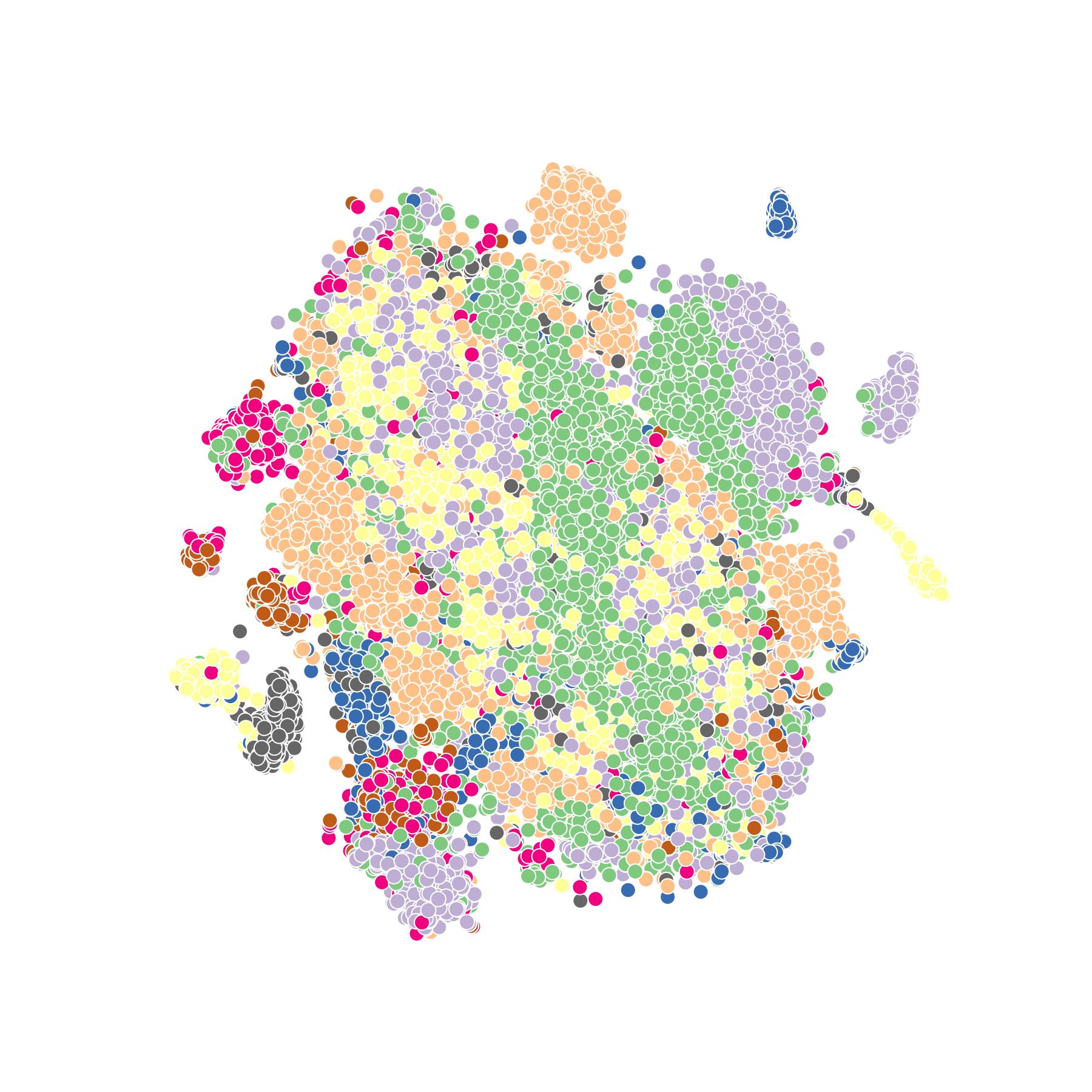}}
    \label{fig_s1}
    \hfil
    \subfigure[]{\includegraphics[width=1.5in]{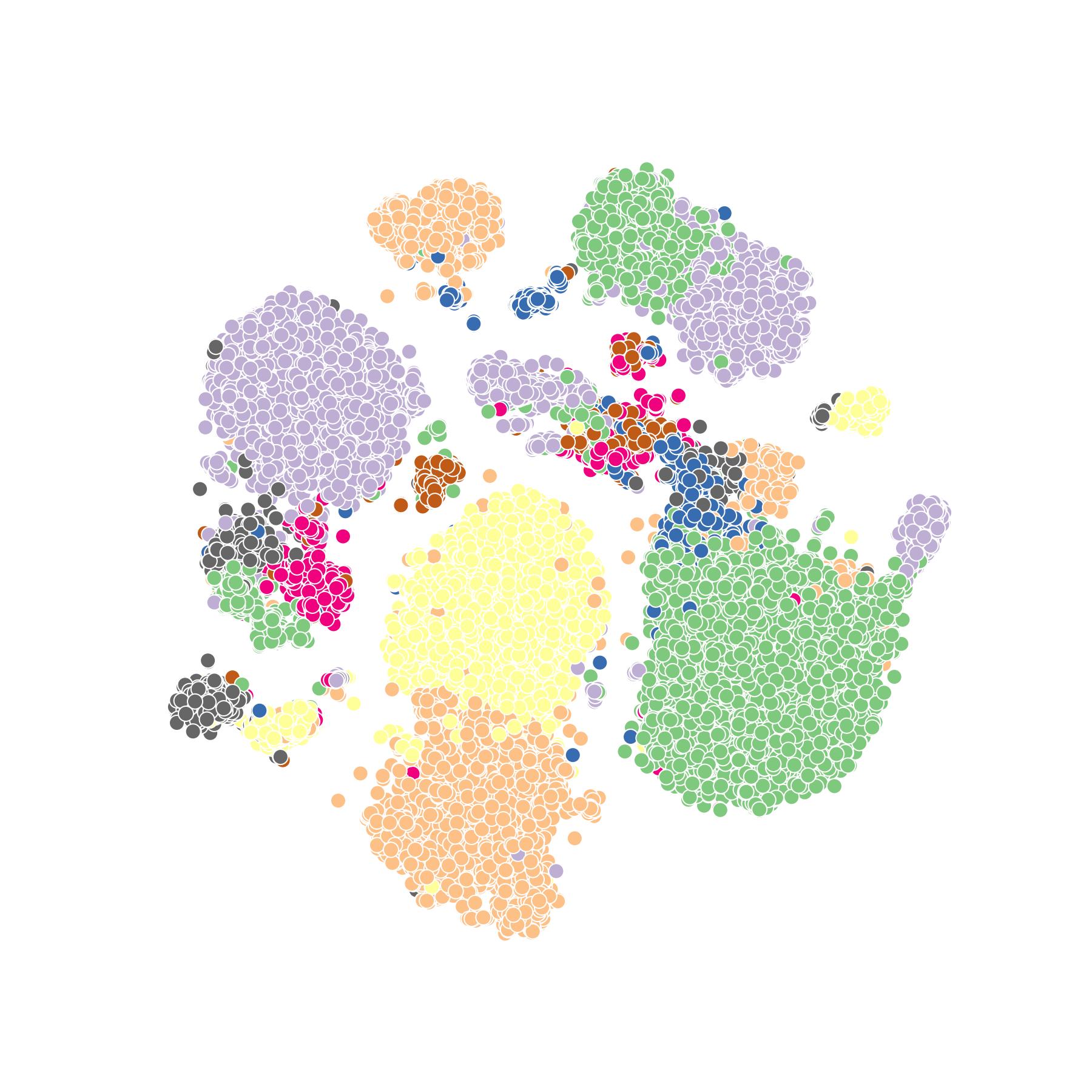}}
    \label{fig_s2}
    \hfil
    \subfigure[]{\includegraphics[width=1.5in]{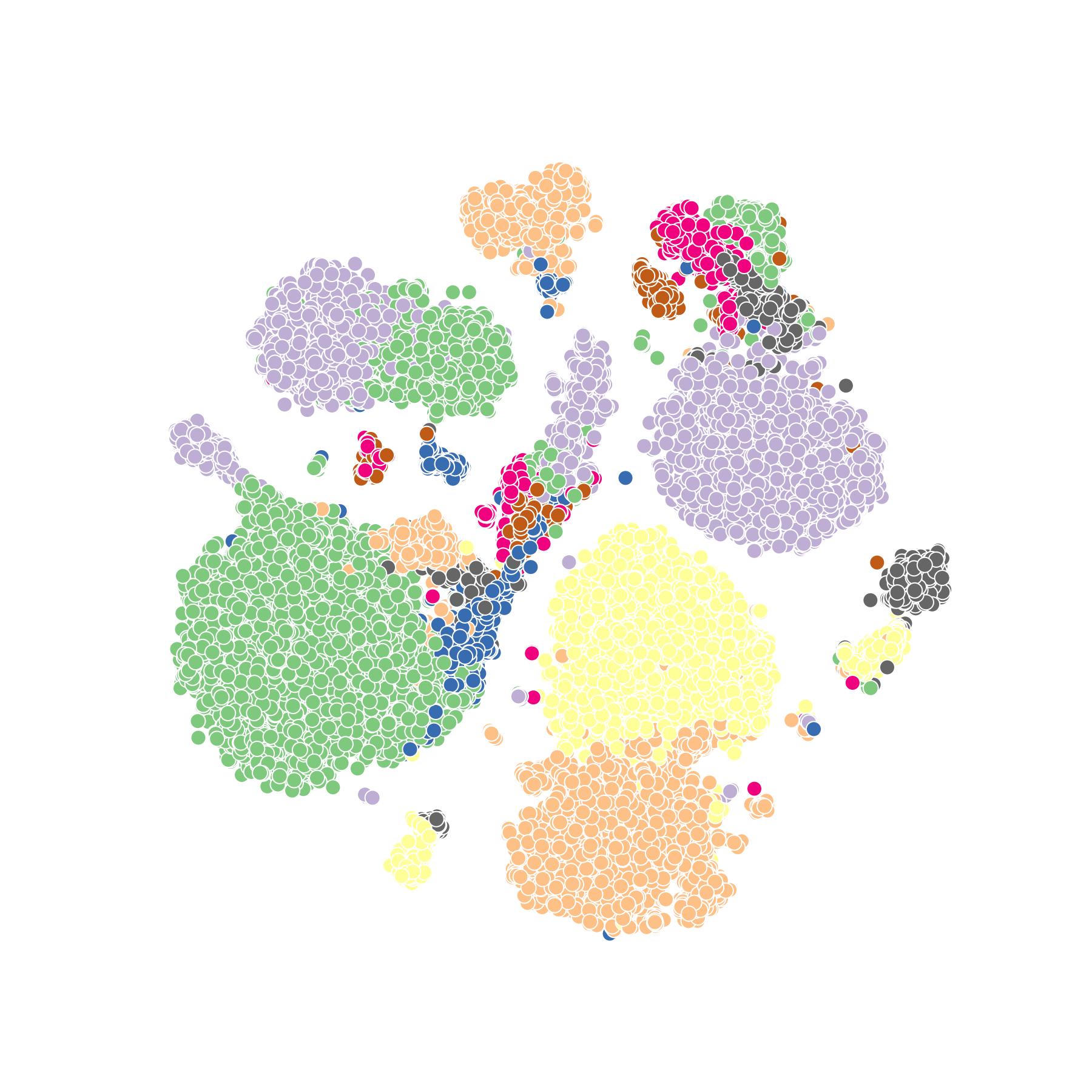}}
    \label{fig_fs3}
    \hfil
    \subfigure[]{\includegraphics[width=1.5in]{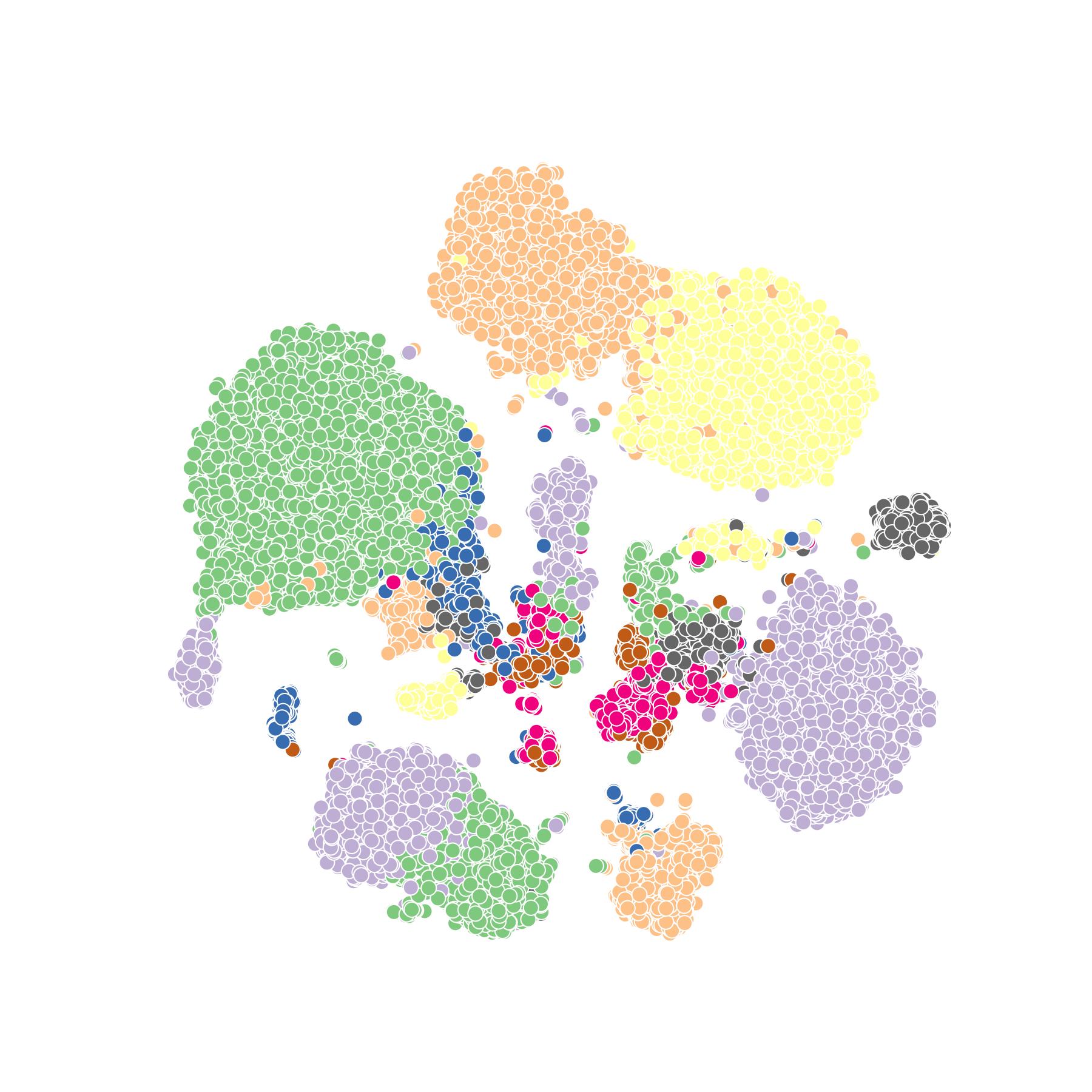}}
    \label{fig_s4}
    
    \caption{Visual representation showcasing the distinct embeddings in the absence of various modules, on the BMNC dataset. (a) scUNC excluding the reconstruction loss. (b) scUNC excluding the clustering loss. (c) scUNC excluding the CVFN module. (d) scUNC with all modules retained.}
\label{fig4}
\end{figure*}

\subsubsection{Implementation Details}

The experiments were conducted on a personal workstation running the Linux operating system with the following specifications: i9-12900KF CPU, 64GB of RAM, and a GeForce RTX 3070Ti GPU. The autoencoder structure utilized in this framework consists of four layers with the following dimensions: (512, 256, 128, 100). A learning rate of 0.0001 was configured, and the batch size was set to 1024. All experimental performance results were obtained using the optimal parameters. The demo code of our work can be accessed publicly at the following link \footnote{https://github.com/DayuHuu/scUNC}.

\subsubsection{Evaluation}

This work employed four widely-used metrics for cluster evaluation, namely, adjusted Rand index (ARI), normalized mutual information (NMI), accuracy (ACC), and purity (PUR), to assess clustering performance. Given their widespread use, here, we provide a brief introduction and formula definitions.

\begin{itemize}
\item \textbf{ARI} takes into account the possibility of random cluster assignment and is defined as:
\begin{equation}
\label{eq:15}
\begin{aligned}
\text{ARI} = \frac{\sum_{ij}\binom{n_{ij}}{2}-[\sum_{i}\binom{a_{i}}{2}\sum_{j}\binom{b_{j}}{2}]/\binom{n}{2}}{\frac{1}{2}[\sum_{i}\binom{a_{i}}{2}+\sum_{j}\binom{b_{j}}{2}]-[\sum_{i}\binom{a_{i}}{2}\sum_{j}\binom{b_{j}}{2}]/\binom{n}{2}}.
\end{aligned}
\end{equation}

\item \textbf{NMI} measures the mutual information between the clustering results and reference labels, and its formula is:
\begin{equation}
\label{eq:16}
\begin{aligned}
\text{NMI}=\frac{2 M I(U, V)}{H(U)+H(V)}.
\end{aligned}
\end{equation}

\item \textbf{ACC} measures the consistency between cluster assignments and reference labels, with the calculation formula as follows:
\begin{equation}
\label{eq:17}
\begin{aligned}
\text{ACC} = \frac{TP + TN}{TP+TN+FP+FN}.
\end{aligned}
\end{equation}

\item \textbf{PUR} measures the proportion of samples within the same cluster that share the same true category and is defined as:
\begin{equation}
\label{eq:18}
\begin{aligned}
\text{PUR}=\frac{1}{N} \sum_{k} \max _{j}\left|w_{k} \cap c_{j}\right|.
\end{aligned}
\end{equation}
\end{itemize}

\subsection{Clustering performance}

Table \ref{table2} presents the performance comparison between scUNC and baseline methods. In each metric, the top two performers are highlighted in bold. From the results, it is evident that scUNC surpasses the other nine comparison methods consistently. In the 12 evaluations conducted, scUNC achieved first place in 8 of them and ranked within the top two in 11 of them. There was a slight decrease in the PUR metric when evaluated on the SMAGE-3K dataset, which is reasonable as no single algorithm can be universally applicable. Analyzing the potential reasons for this phenomenon, we speculate that it may be due to class imbalance, which can lead to the Purity metric overestimating the performance of certain clustering algorithms. For example, if all samples are assigned to the same cluster, the Purity value would be 1, but this does not reflect good clustering performance.

Furthermore, Fig. \ref{fig4} illustrates the two-dimensional visualizations of embeddings generated by removing various modules. It is evident that scUNC with all modules retained exhibits commendable dispersion. In contrast, embeddings generated by models with removed modules showcase some degree of cluster confusion, particularly when the reconstruction loss is excluded. In addition, in Fig. \ref{fig4}(b) and Fig. \ref{fig4}(c) that denotes the cell representations without the clustering loss and CVFN module, we observe the yellow and black clusters at the edges of the figure closely sticking together without dispersion. However, in Fig. \ref{fig4}(d), these two clusters are distinctly separated. Overall, the evaluation of these metrics explicitly demonstrates the substantial enhancement in the quality of embeddings by proposed distinct modules.

\begin{table}[t]
\centering
\renewcommand{\arraystretch}{1.5}
\caption{Ablation study of optimization modules. The top performer is highlighted in bold.}
\vspace{-0.2 cm}
\resizebox{.95\columnwidth}{!}{
\small

\begin{tabular}{ccccc} 
\hline
Datasets                     & Module      & ARI & NMI  & ACC \\ 
\hline
\multirow{3}{*}{BMNC} 
 &Without CVFN &  0.8143	&0.8357&	0.8058\\
& Without Clustering Loss & 0.6844 & 0.7997 & 0.7211\\
& Without Reconstruction Loss& 0.0333 & 0.0881 & 0.2006 \\
& scUNC & \textbf{0.8406} & \textbf{0.8632} & \textbf{0.8530} \\
\hline

\multirow{3}{*}{SMAGE-10K} 
 &Without CVFN & 0.2134 & 0.5186 & 0.3311 \\
& Without Clustering Loss & 0.3442 & 0.5429 & 0.4569\\
& Without Reconstruction  Loss       & 0.1186 & 0.2426 & 0.4517\\
&scUNC & \textbf{0.6067} & \textbf{0.6422} & \textbf{0.6516} \\
\hline

\multirow{3}{*}{SMAGE-3K} 
 &Without CVFN &   0.1481 & 0.4593 & 0.2692 \\
& Without Clustering Loss &  0.379 & 0.5606 & 0.4963 \\
& Without Reconstruction  Loss   & 0.4396 & 0.5292 & 0.6596\\
& scUNC & \textbf{0.5672} & \textbf{0.609} & \textbf{0.6839} \\
\hline

\end{tabular}}
\vspace{-0.2 cm}
\label{table4}

\end{table}

\begin{table}[t]
\centering
\renewcommand{\arraystretch}{1.5}
\caption{Ablation study of merging module. The top performer is highlighted in bold.}
\vspace{-0.2 cm}
\resizebox{.95\columnwidth}{!}{
\small
\begin{tabular}{ccccc} 
\hline
Datasets                     & Module      & ARI & NMI  & ACC \\ 
\hline
\multirow{2}{*}{BMNC} 
 &Without Merging &  0.4560	&0.7544&	0.5062\\
& scUNC & \textbf{0.8406} & \textbf{0.8632} & \textbf{0.8530} \\
\hline

\multirow{2}{*}{SMAGE-10K} 
 &Without Merging  & 0.1631 & 0.4808 & 0.3015 \\
&scUNC & \textbf{0.6067} & \textbf{0.6422} & \textbf{0.6516} \\
\hline

\multirow{2}{*}{SMAGE-3K} 
 &Without Merging  &   0.2049 & 0.4510 & 0.3041 \\
& scUNC & \textbf{0.5672} & \textbf{0.609} & \textbf{0.6839} \\
\hline
                          
\end{tabular}}
\vspace{-0.2 cm}
\label{table5}

\end{table}

\subsection{Ablation Study}
To validate the effectiveness of the modules proposed in our model, we conducted ablation experiments on two sets of model variants to determine the contributions of each module to the overall results.

Specifically, we divided the ablation experiments into two groups, with the first group aimed at verifying the effectiveness of the proposed optimization modules. In this group, we provided three sets of variants, each removing one of the following: the CVFN network, the clustering loss, or the reconstruction loss. We compared the performance of these three variants with the complete scUNC model using previous clustering metrics, as presented in Table \ref{table4}, with the best performance in each metric highlighted in bold. It is evident from the results in the table that all three modules we constructed make a significant contribution to the model, and removing any of them leads to a decrease in overall performance. This indicates that the CVFN network effectively addresses the information richness disparity between scRNA and scATAC data, and the clustering loss and reconstruction loss designed by us play a crucial role in optimizing the model.

The second group of ablation experiments aimed to verify the performance enhancement provided by the automatic merging. In this group, we only provided one set of variants, which removed the merging module we designed. From Table \ref{table5}, we observed a significant drop in model performance, indicating that merging clusters using the dip-test significantly improved clustering performance. In summary, through comprehensive ablation experiments, all modules of the scUNC model have been demonstrated to be effective.

\subsection{Model Analysis}

\subsubsection{Parameter Analysis}
\begin{figure}[t]
  \centering
  \subfigure[PUR] {\includegraphics[width=4.3cm]{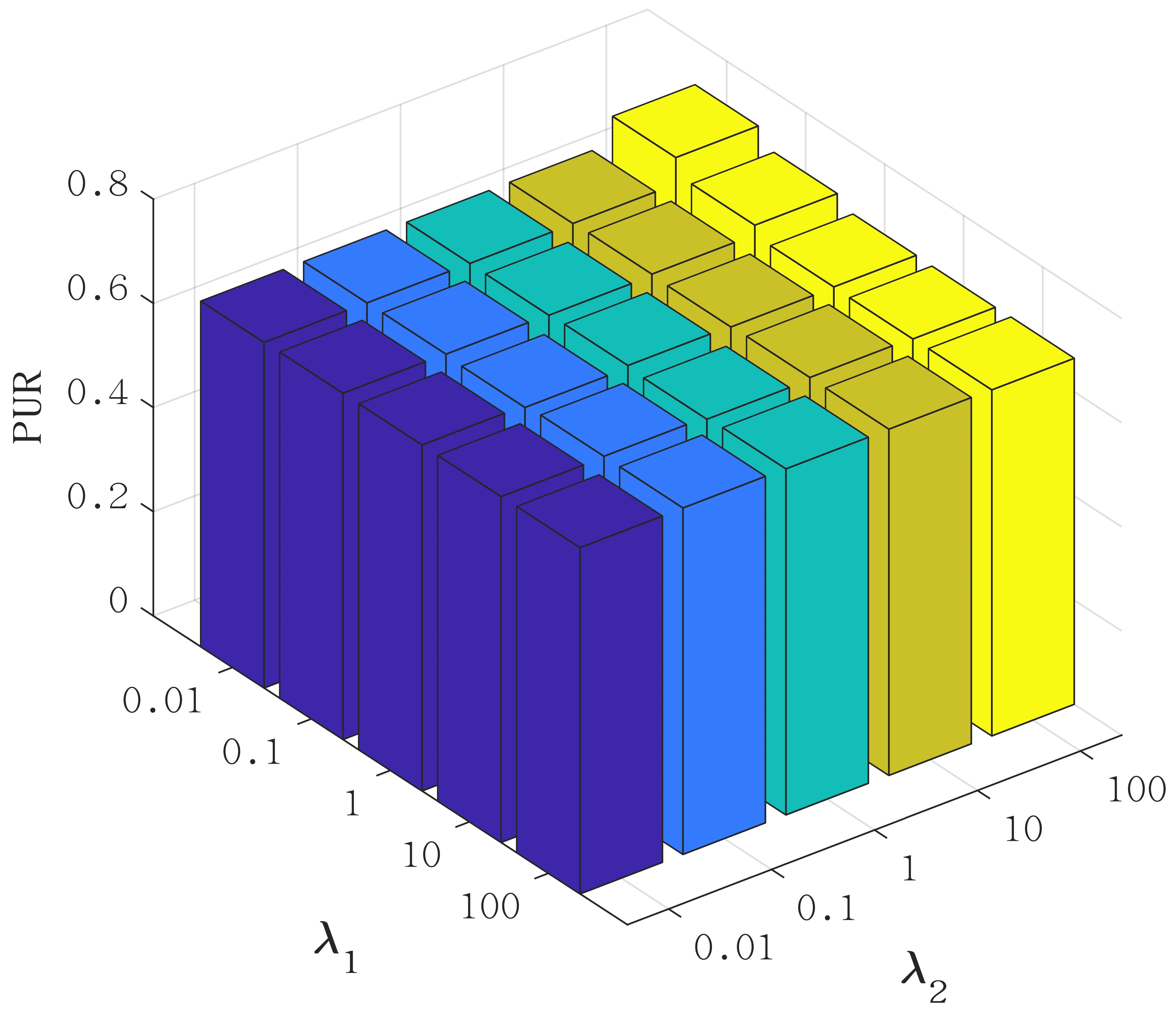}}
  \subfigure[NMI] {\includegraphics[width=4.4cm]{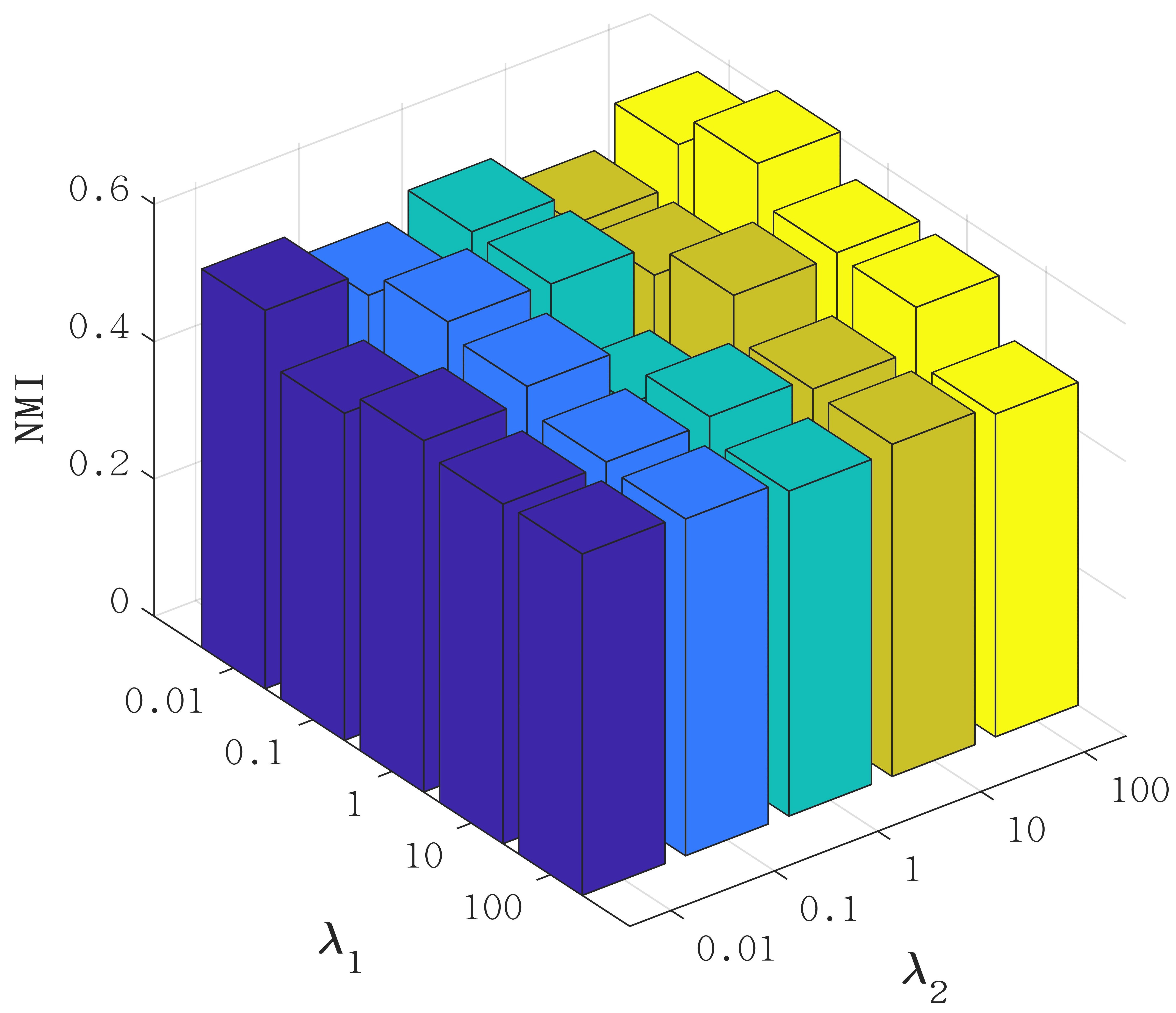}}
\caption{Investigation of hyperparameter $\lambda_1$ and $\lambda_2$ by PUR and NMI.}
\label{fig5}
\end{figure}

In the workflow of this study, we introduced two hyperparameters to balance the clustering loss and reconstruction loss during optimization. In this section, we conducted sensitivity experiments on these parameters using the SMAGE-3K dataset, thoroughly analyzing various values for both hyperparameters. As shown in Fig. \ref{fig5}, we visually represented the clustering performance under different values of $\lambda_1$ and $\lambda_2$ through three-dimensional visualization.

Fig. \ref{fig5}(a) shows the evaluation based on the PUR metric, exhibiting minimal fluctuations in its results. This signifies the stability of our algorithm's performance, with the two hyperparameters having limited impact on the PUR metric, displaying a low sensitivity. However, upon examining Fig. \ref{fig5}(b), we observe smaller variations in performance, indicating that the hyperparameter settings do influence the model's NMI to some extent. Notably, we discovered that setting $\lambda_1$ to 0.1 and $\lambda_2$ to 100 yields the model's optimal performance. Therefore, we configure the two hyperparameters accordingly to achieve the best performance of the model.






\subsubsection{Convergence analysis}
\begin{figure*}[!t]
\centering
\subfigure[BMNC]{\includegraphics[width=3in]{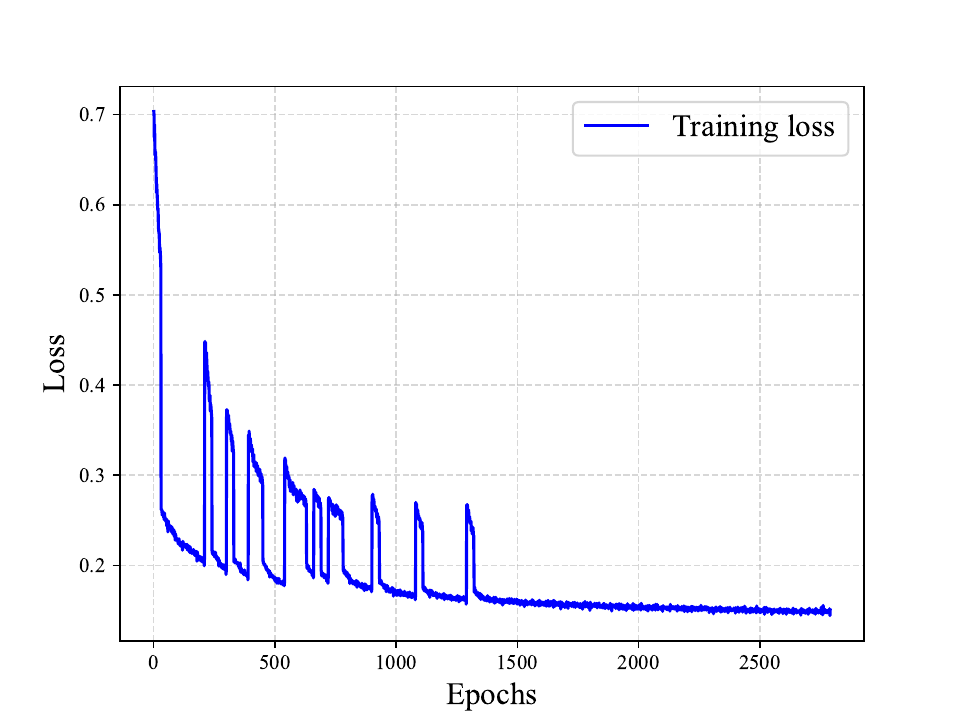}
\label{fig_1}}
\hfil
\subfigure[SMAGE-10K]{\includegraphics[width=3in]{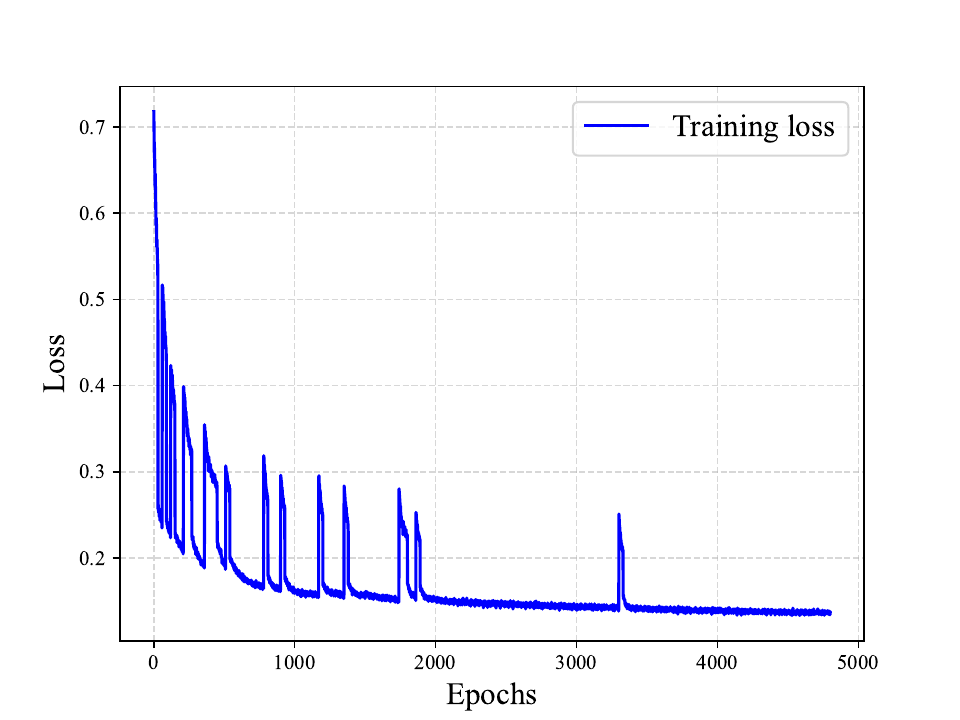}
\label{fig_2}}

\subfigure[SMAGE-3K]{\includegraphics[width=3in]{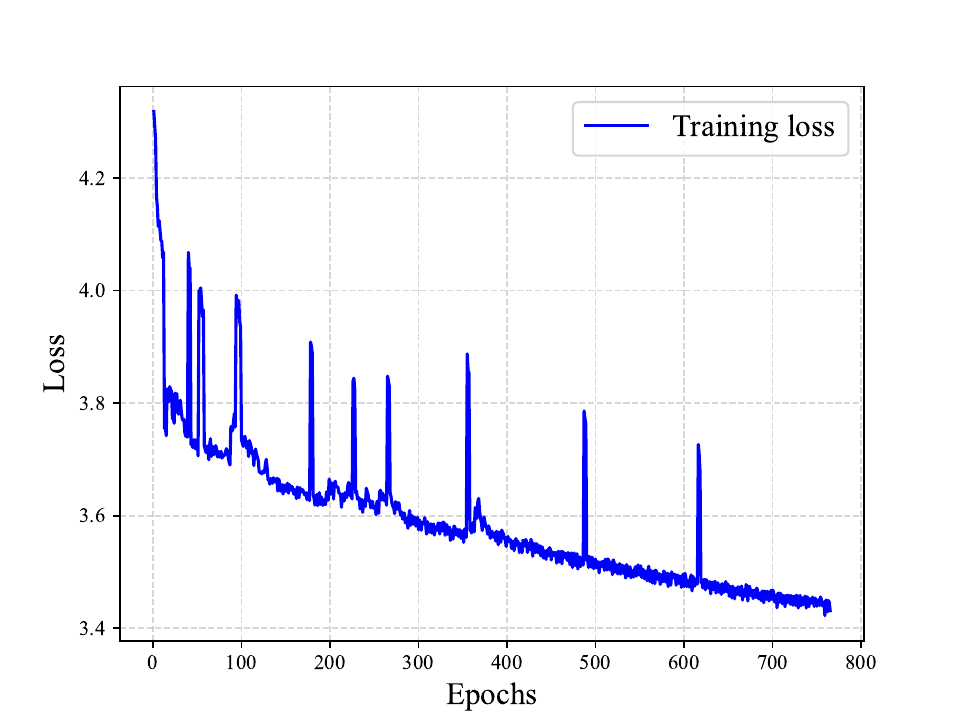}
\label{fig_3}}
\hfil
\subfigure[WikipediaArticles]{\includegraphics[width=3in]{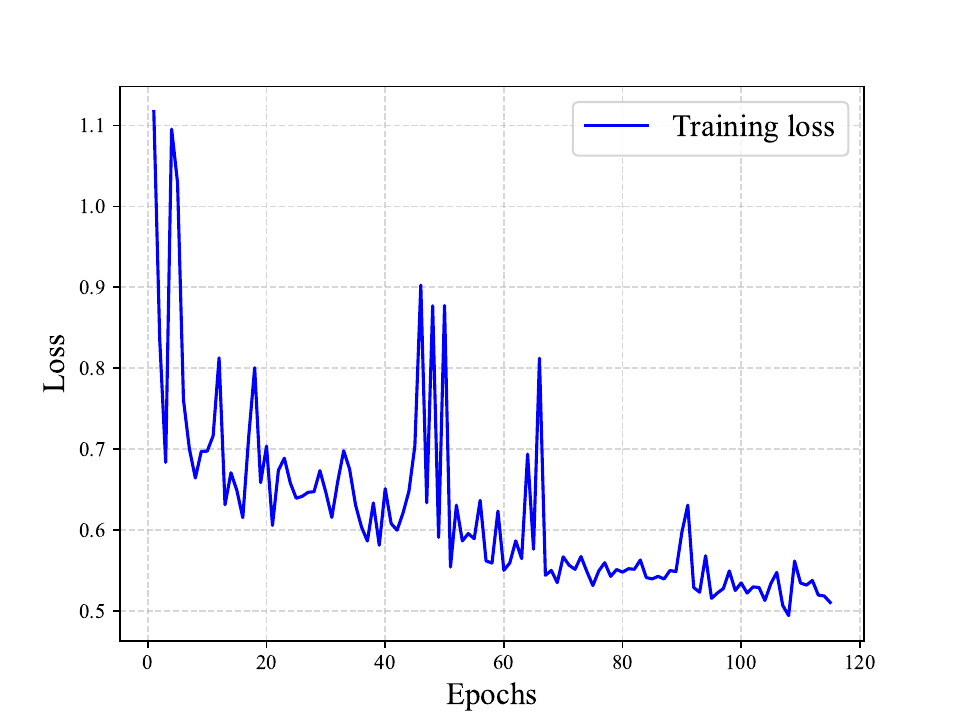}
\label{fig_4}}

\caption{The optimization process of the objective function on four benchmark datasets. (a) BMNC. (b) SMAGE-10K. (c) SMAGE-3K. (d) WikipediaArticles.}
\label{fig6}
\end{figure*}

To illustrate the convergence of the proposed scUNC, the objective values of the loss function are plotted in Fig. \ref{fig6}. From this figure, we can see that the values decrease until convergence. It is important to note that there are some significant fluctuations in the curve. This phenomenon arises due to the iterative merging and optimization of clusters within our algorithm. As a result, the loss value is recalculated after each cluster merging, leading to fluctuations. However, from the results depicted in the figure, it is evident that the fluctuations in the loss value decrease continuously, eventually settling into narrow fluctuations. These findings provide further confirmation of the convergence of scUNC.

\subsection{Generalization analysis}
The Generalization Experiment is conducted to assess the model's capability on unseen data. Here, we explore the generalizing capability of scUNC approach designed for single-cell data on non-cellular data.

The specific details of the non-cellular multi-view dataset have previously been presented in Table \ref{table3}. Here, our competing methods remain the clustering algorithms specifically designed for single-cell data. The evaluation results, as shown in Table \ref{table6}, clearly demonstrate that our scUNC algorithm, while tailored for single-cell multi-view data, still achieves excellent performance on non-cellular multi-view data. This indicates that our model possesses strong generalization capabilities and has the potential to extend to more scenarios.

In contrast, most clustering methods specialized for single-cell data experience significant performance degradation on non-cellular datasets. Taking the scMDC method as an example, as presented in Table \ref{table2}, it performs well when dealing with single-cell data. However, when applied to non-cellular datasets, its performance deteriorates significantly. This suggests that most single-cell clustering algorithms lack the ability to generalize to other scenarios. In summary, we have demonstrated that our model not only performs exceptionally but also exhibits strong generalization capabilities. In the future, it has the potential to be extended into a versatile multi-view clustering framework.

\begin{table}[t]
\centering
\renewcommand{\arraystretch}{1.5}
\caption{Generalization results on WikipediaArticles dataset. The top two performers is highlighted in bold.}
\vspace{-0.2 cm}
\small
\begin{tabular}{c|c|c|c|c}
\hline
Datasets & \multicolumn{4}{c}{WikipediaArticles} \\
\hline
Metrics & ARI & NMI & PUR & ACC  \\
\hline
$k$-means & 0.4204 & 0.5342 & 0.5628 & \textbf{0.6089}  \\
Spectral &0.2721 & 0.5250 & 0.5584 & 0.5036 \\
DESC &   \textbf{0.4359} & 0.5521 & \textbf{0.6349} & 0.6017 \\
scDeepCluster & 0.3846 & 0.5520 & 0.5830 & 0.5310 \\
scDSC   & 0.3634 & \textbf{0.5572} & 0.5729 & 0.5541 \\
DCCA &0.2953  & 0.4177 & 0.5209  & 0.5123 \\
scMVAE & 0.0391 & 0.1031 & 0.2092  & 0.2439 \\
scMCs  & 0.0205 & 0.0574& 0.2266 & 0.1962 \\
scMDC  & 0.3201 & 0.4639 & 0.5382 & 0.4834  \\
scUNC (Ours)  & \textbf{0.4563} & \textbf{0.5539} & \textbf{0.6176} & \textbf{0.6046}  \\
\hline
\end{tabular}
\vspace{-0.2 cm}
\label{table6}

\end{table}

\section{Conclusion}

In this paper, we present a pioneering No-K MVC framework tailored for single-cell data. By employing the CVFN network, we effectively address the issue of the unbalanced information richness between different cell views. Furthermore, to overcome the limitation of manually determining the number of clusters, we introduce an automatic clustering module that leverages community detection for creating initial clusters and employs dip-test detection iteratively for cluster merging until convergence. This module not only enhances the quality of clusters but also offers a user-friendly approach for medical practitioners engaged in cell analysis, as it eliminates the need for manual cluster number specification. Extensive experimental results demonstrate that our method achieves SOTA performance in both single-cell and non-cellular data without a predefined number of clusters.

Furthermore, there are still some limitations here. Firstly, due to the early stage of multi-modal single-cell sequencing, publicly available multi-view datasets are not easily accessible, which hinders a comprehensive evaluation of the proposed model. Secondly, existing methods for defining neighbors still rely on KNN, overlooking the potential higher-order neighborhood information that may exist among cells. Nonetheless, we have presented a viable approach for the automated identification of communities among cells. In the future, we plan to further analyze the effectiveness of each view and extend the neighborhood information to higher orders to uncover higher-order community relationships among cells.

\section*{Acknowledgments}
This work was supported in part by the National Key R\&D Program of China (no. 2020AAA0107100), and the National Natural Science Foundation of China (no. 62325604, 62276271).

\bibliographystyle{IEEEtran}
\bibliography{IEEEabrv,ref}

\begin{IEEEbiography}[{\includegraphics[width=1in,height=1.10in,clip,keepaspectratio]{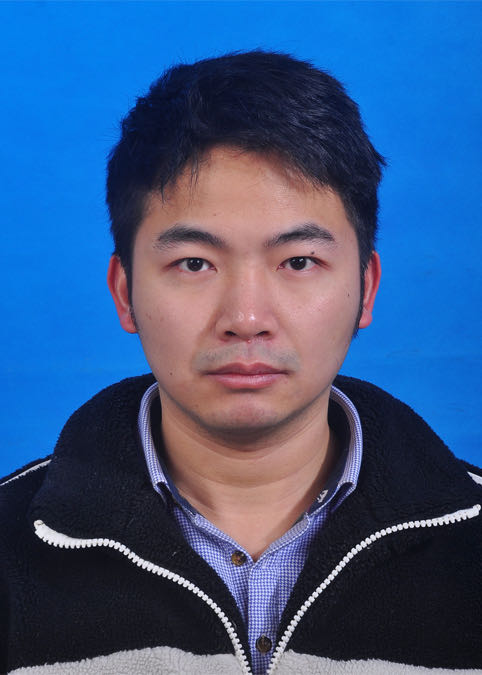}}]
{Dayu Hu} is currently pursuing a Ph.D. degree at the National University of Defense Technology (NUDT). Before joining NUDT, he got his BSc degree at Northeastern University (NEU). His current research interests include graph learning and bioinformatics. He has published several papers and served as PC member/ Reviewer in highly regarded journals and conferences such as ACM MM, AAAI, TNNLS, TKDE, TCBB, etc. 
\end{IEEEbiography}

\begin{IEEEbiography}[{\includegraphics[width=1in,height=1.10in,clip,keepaspectratio]{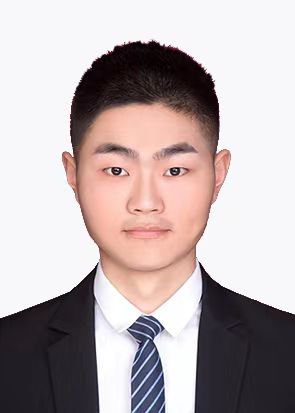}}]
{Zhibin Dong} is currently pursuing the Ph.D. degree
with the National University of Defense Technology(NUDT), Changsha, China. He has published several papers and served as
a Program Committee (PC) member or a reviewer for top conferences, such as IEEE Conference on Computer Vision and Pattern Recognition (CVPR), IEEE International Conference on Computer Vision(ICCV), Association for Computing Machinery’s
Multimedia Conference (ACM MM), Association for the Advancement of Artificial Intelligence Conference (AAAI), and the IEEE TRANSACTIONS ON NEURAL NETWORKS AND LEARNING SYSTEMS (TNNLS). His current research interests include graph representation learning, deep unsupervised learning, and multi-view clustering.
\end{IEEEbiography}

\begin{IEEEbiography}[{\includegraphics[width=1in,height=1.10in,clip,keepaspectratio]{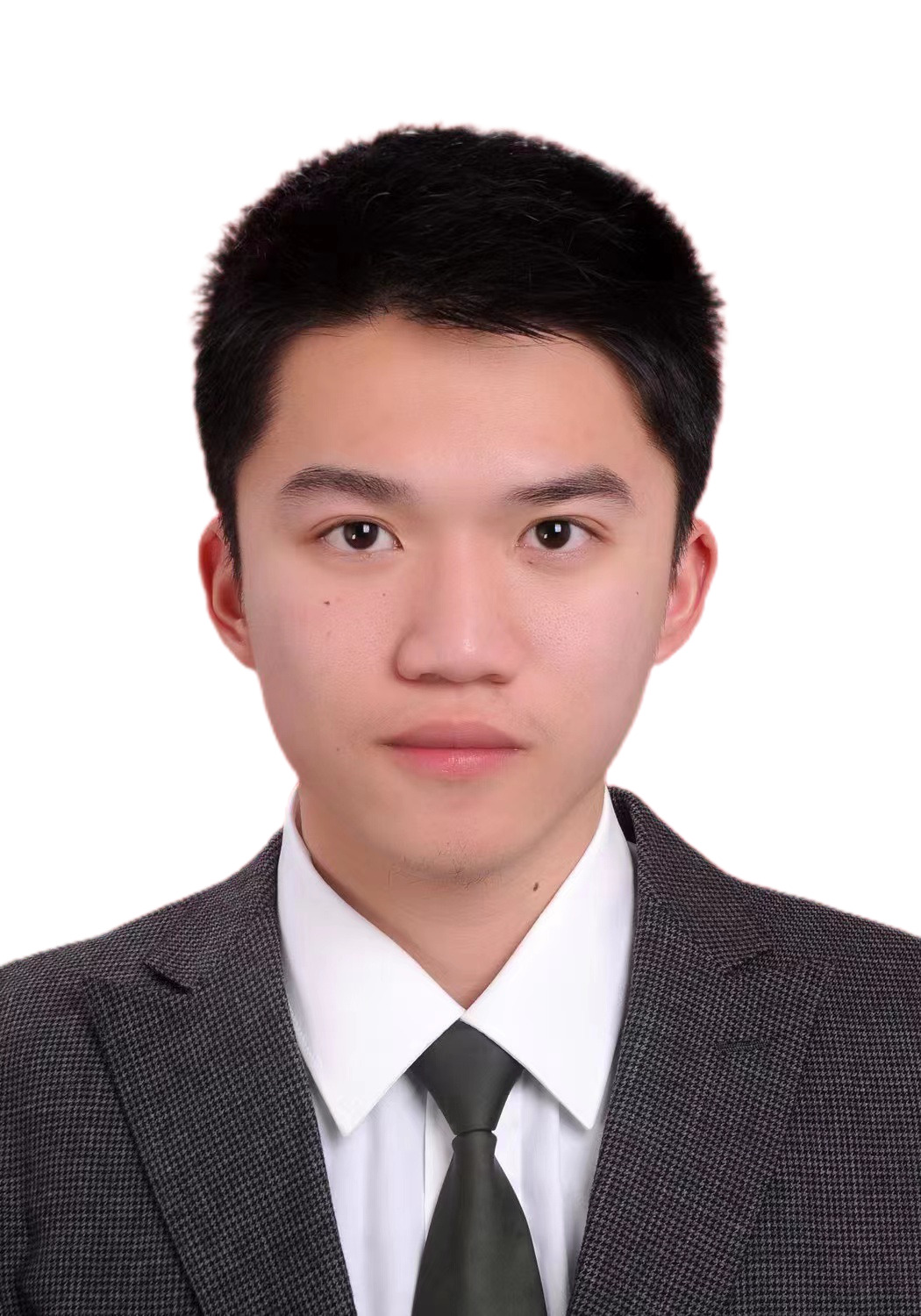}}]
{Ke Liang} is currently pursuing a Ph.D. degree at the National University of Defense Technology (NUDT). Before joining NUDT, he got his BSc degree at Beihang University (BUAA) and received his MSc degree from the Pennsylvania State University (PSU). His current research interests include knowledge graphs, graph learning, and healthcare AI. He has published several papers in highly regarded journals and conferences such as SIGIR, AAAI, ICML, ACM MM, IEEE TNNLS, IEEE TKDE, etc.
\end{IEEEbiography}

\begin{IEEEbiography}[{\includegraphics[width=1in,height=1.10in,clip,keepaspectratio]{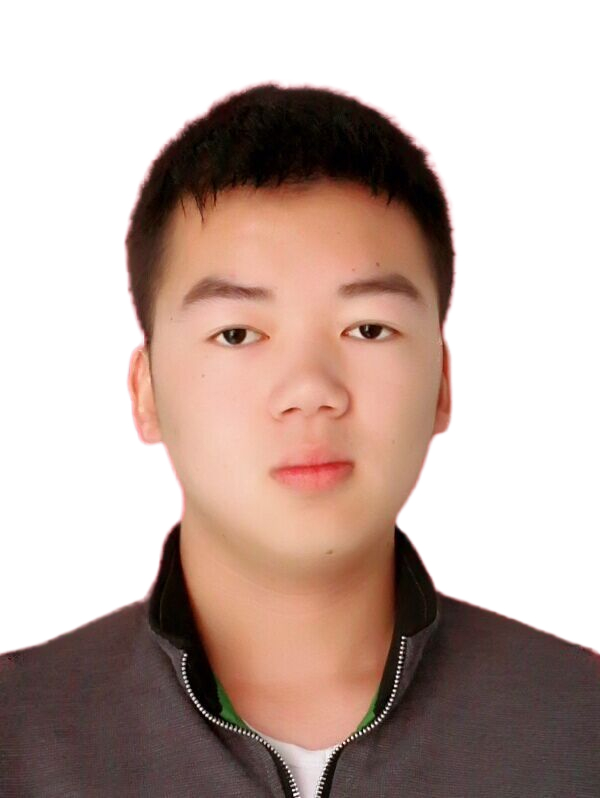}}]
{Jun Wang} received the MS degree from China University of Geosciences, Wuhan, China, in 2023. Currently, he is pursuing his Ph.D. degree at School of Computer Science, National University of Defense Technology. His research interest is multi-view learning.
\end{IEEEbiography}

\begin{IEEEbiography}[{\includegraphics[width=1in,height=1.25in,clip,keepaspectratio]{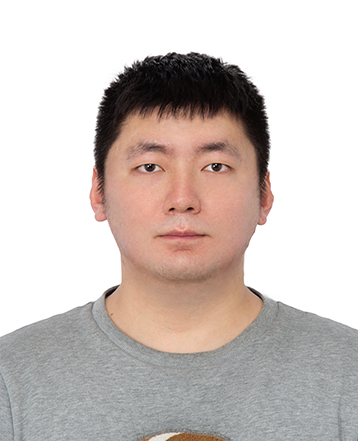}}]{Siwei Wang} is currently assistant research professor with the N Intelligent Game and Decision Lab, China. He has published several papers and served as a AC/SPC /Reviewer in top journals and conferences, such as IEEE TRANSACTIONS ON PATTERN ANALYSIS AND MACHINE INTELLIGENCE(TPAMI), IEEE TRANSACTIONS ON KNOWLEDGE AND DATA ENGINEERING (TKDE), IEEE TRANSACTIONS ON NEURAL NETWORKS AND LEARNING SYSTEMS (TNNLS), IEEE TRANSACTIONS ON IMAGE PROCESSING (TIP), IEEE TRANSACTIONS ON CYBERNETICS (TCYB), IEEE TRANSACTIONS ON MULTIMEDIA (TMM), ICML, NeurIPS, CVPR, ICCV, ECCV, AAAI, and IJCAI. His current research interests include multi-modal, unsupervised multi-view learning, scalable clustering, and deep unsupervised learning.
\end{IEEEbiography}

\begin{IEEEbiography}[{\includegraphics[width=1in,height=1.10in,clip,keepaspectratio]{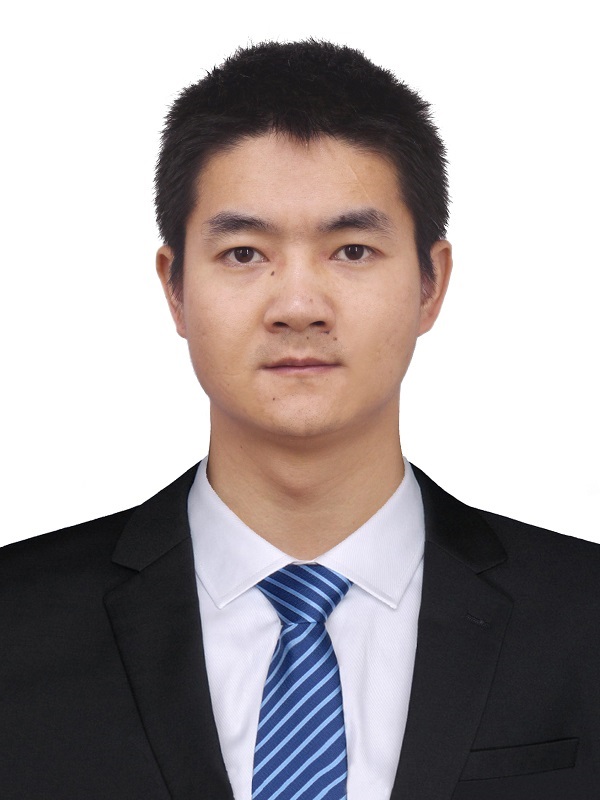}}]{Xinwang Liu} received his PhD degree from National University of Defense Technology (NUDT), China. He is now Professor of School of Computer, NUDT. His current research interests include kernel learning and unsupervised feature learning. Dr. Liu has published 60+ peer-reviewed papers, including those in highly regarded journals and conferences such as IEEE T-PAMI, IEEE T-KDE, IEEE T-IP, IEEE T-NNLS, IEEE T-MM, IEEE T-IFS, ICML, NeurIPS, ICCV, CVPR, AAAI, IJCAI, etc. He serves as the associated editor of TNNLS, TCYB and Information Fusion Journal. More information can be found at {https://xinwangliu.github.io/}.
\end{IEEEbiography}

\end{document}